%% file: top-arxiv.tex
\documentclass[10pt,pdftex,a4paper]{article}

\usepackage[utf8x]{inputenc}
\usepackage{amsmath}
\usepackage{amsfonts}
\usepackage{amsthm}
\usepackage{amssymb}
\usepackage{stmaryrd}
\usepackage{mathabx}
\usepackage{rotating}
\usepackage{ppprog}       %
\usepackage{shorthand}    %
\usepackage{veralgs}      %
\usepackage{caption}
\usepackage{subfig}
\usepackage[OT1]{fontenc}
\usepackage{paralist}
\usepackage{hhline}
\usepackage{tabularx}

\usepackage{listings}
\usepackage{pxfonts}
\usepackage{xcolor}
\definecolor{javared}{rgb}{0.8,0.1,0.1} %
\definecolor{javablue}{rgb}{0.2,0.2,0.8} %
\definecolor{javagreen}{rgb}{0.1,0.6,0.1} %
\definecolor{javapurple}{rgb}{0.5,0,0.35} %
\definecolor{javadocblue}{rgb}{0.25,0.35,0.75} %
\usepackage{xcolor}
\usepackage{framed}
\colorlet{shadecolor}{yellow!20}
\usepackage[a4paper, total={7in, 8.5in}]{geometry}

\def\benchmark{Helenos}%

\begin{document}

\title{\benchmark: A Realistic Benchmark for Distributed Transactional Memory}
\date{Mar 21, 2016}
\author{Jan Baranowski,
    Pawe\l{} Kobyli\'nski,
    Konrad Siek,
    Pawe\l{} T. Wojciechowski \\
Institute of Computing Science\\
Pozna\'n University of Technology
}

\maketitle

\begin{abstract}

\input{abstract}

\end{abstract}

\input{introduction}

\input{relatedwork}

\input{fb}

\input{evaluation}

\input{conclusions}

\paragraph{Acknowledements}

The project was funded from National Science Centre funds granted by decision
No. DEC-2012/07/B/ST6/01230.

\bibliographystyle{abbrv}
\bibliography{top-arxiv}

\end{document}

%% file: abstract.tex
Transactional Memory (TM) is an approach to concurrency control that aims to make writing parallel programs both effective and simple. The approach is started in non-distributed multiprocessor systems, but is gaining popularity in distributed systems to synchronize tasks at large scales.  Efficiency and scalability are often the key issues in TM research, so performance benchmarks are an important part of it. However, while standard TM benchmarks like the STAMP suite and STMBench7 are available and widely accepted, they do not translate well into distributed systems. Hence, the set of benchmarks usable with distributed TM systems is very limited, and must be padded with microbenchmarks, whose simplicity and artificial nature often makes them uninformative or misleading. Therefore, this paper introduces Helenos, a realistic, complex, and comprehensive distributed TM benchmark based on the problem of the Facebook inbox, an application of the Cassandra distributed store.

%% file: introduction.tex
\section{Introduction}
\label{sec:introduction}

\emph{Transactional Memory (TM)} \cite{HM93} is an approach to concurrency
control that aims to make writing parallel programs both effective and simple.
The programmer applies the \emph{transaction} abstraction to denote
sections of code whose atomicity must be preserved. The TM system is then
responsible for enacting the required guarantees, and doing so efficiently.
Efficiency is often the key question in TM research, and various TM systems
employ different concurrency control algorithms that can achieve quite
divergent levels of performance depending on the workload. Hence, there is a
need for empirical evaluation of their performance and the trade-offs between
efficiency and features.

Initially, TMs were evaluated using microbenchmarks, but these test specific
features in isolation and use data structures that are too trivial to draw
general conclusions about a TM. Alternatively, there are HPC benchmark suites,
but these are difficult to translate meaningfully into the transactional idiom.
That is, benchmarks from SPEComp \cite{ADE+01} or SPLASH-2 \cite{WOT+95} are
already expertly optimized to minimize synchronization, so any incorporated
transactions are used rarely and have little effect on overall performance. 
Hence, a set of TM-specific benchmarks was needed, whose transactional
characteristics and contention for shared resources were both varied and
controllable. Thus, benchmarks and benchmark suites like STMBench7
\cite{GKV07}, LeeTM \cite{AKJ+08}, and STAMP \cite{MCKO08} were developed.

Given that distributed systems face the same synchronization problems as their
multiprocessor counterparts, distributed transactions are successfully used
wherever requirements for strong consistency meet wide-area distribution, e.g.,
in Google's Percolator \cite{PD10} and Spanner \cite{COR12}. 
Distributed TM adds additional flexibility on top of distributed transactions and allows to lay the groundwork for well-performing
middleware with concurrency control. 
However, in such environments, unlike its non-distributed analogue, distributed
TM must also deal with a flood of additional problems like fault tolerance and
scalability in the face of geo-distribution or heterogeneity.

Distributed TMs were developed by
using transactions on top of replication \cite{BAC08,CRCR09,KKW13,HPR14} or by
allowing clients to atomically access shared resources or to atomically execute
remote code \cite{SR11,TRP13,SW14-hlpp}.
The latter type are of particular interest and the main focus of this paper,
since they incorporate the distributed elements directly into the transaction
abstraction and apply to a wide variety of distributed system models from
cluster computing to clouds and web services.

As with non-distributed TMs, the variety of differently-featured distributed
TMs require empirical evaluation to find how their features, the workloads, and
the configuration of distributed systems influences their performance.
Therefore, as with non-distributed TMs, they must be evaluated empirically.
However, the existing TM benchmarks are not appropriate for distributed TMs.
This is primarily the case
since the structures they use are not easy to distribute.
Distributing non-distributed TM benchmarks often leads to arbitrary sharding of the structure
that has no purpose for the application itself (e.g., the clients still must
access the entire domain). Hence distributing STMBench7, LeeTM, or
\emph{labyrinth} or \emph{k-means} from STAMP creates applications that do not
reflect realistic use cases for distributed systems. On the other hand, even if
a benchmark has valid distributed variants, the conversion is often non-trivial
and should not be expected to be done \emph{ad hoc}, if it is to be uniformly
applied by various research teams.
As a result systems like HyFlow \cite{SR11}, HyFlow2 \cite{TRP13}, and Atomic RMI
\cite{SW14-hlpp} are all evaluated using a few microbenchmarks supplemented by
a distributed version of the \emph{vacation} benchmark from STAMP, which originally
mimics a distributed database use case. In effect, the results of the evaluation are
not always satisfactory because of the simplicity of the testing procedure. There is also
little research showing comparisons between the performance of distributed TMs.
Hence, in this paper we introduce \benchmark{}, a new complex benchmark dedicated for
distributed TMs, that would help to remedy the lack of tools for evaluating such systems. 
The benchmark proposed here is also the first step towards a creation of a comprehensive
framework for evaluating all facets of distributed TMs.

The usefulness of the \benchmark{} benchmark is demonstrated by an analysis of an example
evaluation of four different distributed concurrency control mechanisms and the produced results. 
The aforementioned absence of other benchmarks that can fit a similar role makes conducting a 
straightforward comparison between \benchmark{} and its competitors impossible. Thus, the only 
way for us to showcase \benchmark{} is to perform an example evaluation presented in this paper 
and show that the obtained results are understandable, predictable and explicable, as well as 
meaningful.

This paper is structured as follows. \rsec{sec:rw} discusses the existing TM
benchmarks in more detail and describes how they can be used for evaluating
distributed TMs or what prevents them from being practical for that aim.
\rsec{sec:benchmark} introduces the \benchmark{} benchmark and provides detail
as to its data model, executed transactions, and defined metrics.
Then, in \rsec{sec:evaluation} we evaluate two distributed TMs and two lock-based distributed concurrency control mechanisms.
Finally, we conclude in \rsec{sec:conclusions}.

%% file: relatedwork.tex
\section{Related Work}
\label{sec:rw}

There are a number of TM benchmarks for non-distributed TM systems that are
noteworthy. We concentrate on the ones used more often (STAMP and STMBench7),
but we also present some less known benchmarks further below. We also give our
attention to the benchmarks bundled with HyFlow.

\subsection{STAMP}

The most prevalent suite of benchmarks in use with non-distributed TM is the
STAMP benchmark introduced in \cite{MCKO08} and retrofitted to work with more
modern TM systems in \cite{RLS14}. It consists of eight applications, each
presenting a different algorithm, as a whole providing a wide range of
transaction characteristics. 

Out of the eight benchmarks within STAMP, only some can be used as distributed
applications. 
The best candidate for a distributed benchmark is \emph{vacation}. It simulates
a travel agency application with a database of hotels,
cars, and flights, where a number of clients attempt to book one of each, while
offers are sporadically modified. The database is homogeneous in
nature and can easily be distributed without incurring major modifications on
the benchmark. Hence, the benchmark is often used to evaluate distributed TMs
\cite{SW14-hlpp,MTPR13}. 
However, the benchmark has a limited range of transaction types, so it does not
constitute a comprehensive evaluation tool.

Other benchmarks which lend themselves to distribution are \emph{genome},
\emph{bayes}, \emph{ssca2}, \emph{yada}, and \emph{intruder}. The applications
are based respectively on algorithms for gene sequencing, learning structures
of Bayesian networks, creating efficient graph representations, Delaunay mesh
refinement, and detecting intrusions in a network. The processing in each of
these applications follows a pipeline where a complex data structure is
processed into another form in a series of steps by multiple simultaneous
threads. Such processing can be distributed among several network nodes (e.g.
in high performance clusters) in order to provide more system resources
(processing power, memory) to the algorithms. The issue with producing such
distributed versions of these algorithms is that it cannot be done \emph{ad
hoc}, and often requires that an expert-prepared variant exists 
with at least a reference distributed implementation that could be simply
instrumented by TM researchers. In any case, from the perspective of
distributed system architecture \emph{genome}, \emph{bayes}, \emph{ssca2},
\emph{yada}, and \emph{intruder} are all an example of a single use-case, and
therefore even though these applications provide breadth for a concurrent
benchmark, they provide little breadth in the distributed context.

The \emph{kmeans} benchmark represents a K-means clusterer known from data
mining.  The algorithm executes in rounds: in every round a thread reads the
values of some partition of objects and designates one of them the new center
of the cluster, then all the data objects are reassigned to the closest center.
While the data for the clusterer can be distributed onto multiple nodes, the
resulting variant does not present a genuine distributed use case,  because
execution in rounds induces too high a level of coordination required among
client threads.
We find the remaining \emph{labyrinth} benchmark to be a similar case. There,
transactions operate on a central data structure representing a 3D maze and
attempt to route a path from one point to another using Lee's algorithm.  Since
a path cannot intersect other paths, conflicting writes must be avoided.  Such
an application can be employed for circuit-board design. However, even if the
maze is distributed, as with \emph{kmeans}, the application has no reflection
in distributed system use cases.

\subsection{STMBench7}

The other popular STM benchmark, STMBench7 \cite{GKV07}, is based on an
object-oriented database benchmark.
In STMBench7 clients perform a wide range of transactions on a shared
tree-based database. The tree, called a module, contains three levels of nodes: 
\begin{inparaenum}[\it a)]
    \item \emph{complex assemblies (CAs)}, whose children are either other
    complex assemblies or base assemblies,
    \item \emph{base assemblies (BAs)}, which link to several composite parts,
    or
    \item \emph{composite parts (CPs)}, the leafs of the tree whose payload is
    a document and a graph.
\end{inparaenum}
Each element contains links to its parent as well as children, allowing
bottom-up or top-down traversal. 
Transactions in STMBench7 are either \emph{traversals} (accessing a path from
root to leaf), \emph{queries} (accessing sets of random nodes), or
\emph{structure modifications} (adding or removing parts of a tree). They can
be either long or short, and either read-only or update.
The benchmark simulates a CAD/CAM/CASE application or a multi-user server.

The benchmark can be used for distributed TMs if the tree structure is spread
among several servers in a network. This can be done in one of three ways: 
The benchmark can be distributed per CP, so that each CP is located on
the same network node as its children, but it can be located on a different
server than its parent base assembly, which, in turn, can be located somewhere
different than its own parent complex assembly, etc. 
From the perspective of the client this leads to a system with a flat
structure, where the tree structure from the original application becomes a
simple collection of distributed objects with some objects referring to other
ones. From the distributed TM's point of view the difference between CPs, BAs,
and CAs largely disappears, since their internal workings will be largely
obscured. It is another practical consequence of this, that the difference between
queries and traversals becomes blurred, since the only 
distinguishing feature between them now is how the access sets of transactions
are selected (from other object in traversals rather than from an index in
queries). In effect, this variant creates a very simple benchmark, which
resembles the bank microbenchmark used with distributed TMs, only with a new
transaction type---structure modifications.
In addition, the benchmark can be distributed per CA or BA, i.e., a BA and its
children are located on the same node, but any two BAs or CAs can be located in
different parts of the network. This variant allows groups of objects to be
treated as a single structure by the clients, and, rather than accessing each
object individually, transactions can treat a BA with all its children jointly.
However, from the point of view of a distributed TM, this again becomes a flat
distributed collection of remote objects. In addition to the former variant,
however, the benchmark is further simplified, because the number of
interdependencies between objects is smaller. This is not offset by the
increased complexity of remote objects, which simply means that processing time
of an access increases, but does not have any other effect on transactional
processing.
Finally, the benchmark can be distributed by adding new modules and
distributing per module. However, as the authors themselves note in
\cite{GKV07}, increasing the number of modules (or distributing by module)
isolates transactions from one another, and is therefore not useful for
evaluating TMs.
Neither of the three variants provide a satisfactory tool for evaluating
distributed TMs, since distribution leads to simplification and divorcement
from the original realistic CAM/CAD/CASE use case.

\subsection{HyFlow Benchmarks}

The microbenchmarks and benchmarks included by the authors of HyFlow
\cite{SR11} in their implementation represent what can be considered to be the
best available set of benchmarks for evaluating distributed TM systems. The
suite consists of three microbenchmarks, bank, loan, and distributed hashtable
(DHT), as well as a distributed version of the vacation benchmark from STAMP
(described above).
DHT is a micro-benchmark where each server node acts as a shard of a 
distributed key-value store.  
Transactions atomically perform a number of writes or a number of reads on some
subset of nodes.
The bank benchmark simulates a straightforward distributed application using
the bank metaphor. Each node hosts a number of bank accounts that can be
accessed remotely by clients who perform transfers between accounts or atomic
reads of several accounts.

Finally, the loan benchmark presents a more complex application where the
execution of transactional code is also distributed among several nodes.  Each
server hosts a number of remote objects that allow write and read operations.
Each client transaction atomically executes two reads or two writes on two
objects.  When a read or write method is invoked on a remote object, then it
also executes two reads or writes (respectively) on two other remote objects.
This recursion continues until it reaches a specified depth. 
Hence, the benchmark is characterized by long transactions and high contention,
as well as relatively high network congestion, and is unique in focusing on the
control flow transactional model.

However, while the benchmark suite is able to shed light on the performance of
distributed TMs and includes a distributed TM-specific applications, it lacks
complex benchmarks and therefore does not comprise a comprehensive evaluation
tool.

\subsection{Other TM Benchmarks}

A number of other applications were used to test TM systems apart from the ones
mentioned above. 
Notably, LeeTM \cite{AKJ+08} is an independent implementation of Lee's
algorithm, analogous to the one used in STAMP's labyrinth. The benchmark has
limited use in distributed TM evaluation for the same reasons as its STAMP
counterpart.
EigenBench \cite{HOCB+10} is a comprehensive and highly configurable
microbenchmark for multiprocessor TM evaluation.  One of its interesting
features is that it allows induction of problematic executions (e.g. convoying)
to observe a TM's behavior in their presence. EigenBench uses a number of flat
arrays as its data structure, so it is easy to distribute for use with
distributed TM. However, EigenBench is designed to test individual
characteristics of a TM, and as such does not generate complex workloads
containing heterogenous transactions.
Another interesting application is Atomic Quake \cite{ZGU+09}, a transactional
implementation of a multithreaded game server, which was used to compare the
performance of TM with locking in a complex realistic use case. The benchmark
uses a central server to coordinate remote clients, whose nature prevents the
application from being employed as a truly distributed benchmark.
Finally, in \cite{KKW13} the authors present a custom benchmark application for
replicated transactional memory called Twitter Clone which simulates the
operation of a social networking service. The benchmark is not useful for
non-replicated TM, but the overall idea influenced the benchmark presented in
this paper.

%% file: fb.tex
\section{The \benchmark{} Benchmark}
\label{sec:benchmark}

We introduce the \benchmark{} benchmark which provides a platform for
comprehensively evaluating a distributed TMs for use in non-uniform large scale
distributed systems. Such a benchmark must contain a wide variety of
transaction types and provide a high level of control over the workload that a
TM can be subjected to. This allows thoroughly to test all aspects of a TM, and
observe its performance in diverse environments. However, it is equally
important that a benchmark be evaluated in a realistic setting, i.e., such
where there exists a need for both transactional memory and for distributed
systems. 
The benchmark is implemented in Java, which allows us to interface with some of
the existing distributed TM systems either also implemented in Java \cite{SR11,
SW14-hlpp} or other languages on the Java Virtual Machine \cite{TRP13}. Places
where transactions start and end must be manually marked for each of the tested
libraries allowing great flexibility of measurement.

In order to achieve our goal of providing a realistic use case for a
distributed TM, we base our benchmark on Cassandra \cite{LM10}, a
distributed storage system for large volumes of structured data that emphasizes
decentralization, scalability, and reliability. Cassandra was created to serve
as a storage system for the user-to-user message inbox in Facebook, where it
must be able to withstand large write-heavy throughputs.
The authors specify a data model that is used for Cassandra's intended
application, as well as two typical search procedures: \emph{term search} (find
a message by keywords) and \emph{interaction search} (find all messages
exchanged between two users). 
Cassandra allows to control consistency of its operations by managing which
replicas reply to certain requests. However, the consistency guarantees can be
further extended by introducing the transaction abstraction. Thus, the
Cassandra implementation of the inbox is both an inherently distributed
application, as well as one which can benefit from using the TM.

Hence, we use Cassandra's application as the Facebook inbox as the basis for
our benchmark for distributed TM. We do this by implementing the data model
specified in \cite{LM10} and supplement the original search procedures with a
comprehensive spectrum of pertinent transactions. On the other hand, we remove
features that are not directly \emph{\'a~propos}, including fault tolerance,
local persistence, and partial replication. This makes any evaluation results
simpler to predict and analyze, but does so without loosing the nuance of the
original benchmark.

\subsection{Data Model}
\label{sec:datamodel}
\def\tkw{\mathbb{T}_{\textit kw}}

\begin{figure}[t]
    \begin{center}
    \includegraphics[width=.55\linewidth]{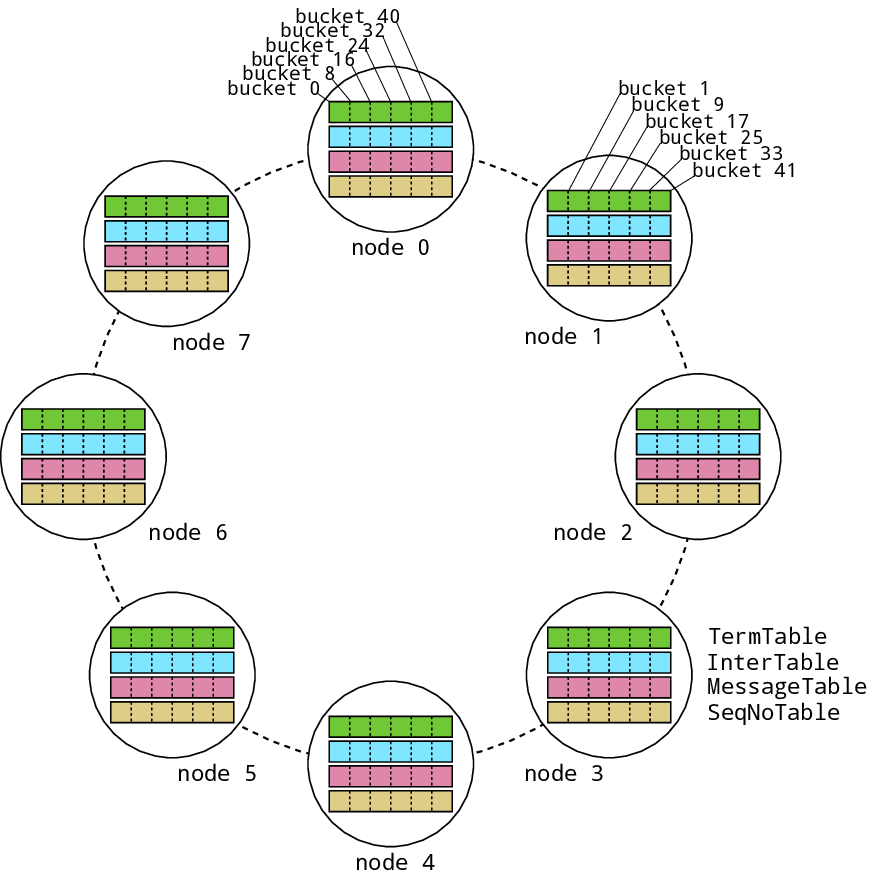}
    \end{center}
    \caption{\label{fig:arch} Database architecture.}
\end{figure}

We follow Cassandra's data model and a logical data model from the inbox
application in \cite{LM10}. 
First of all, a Cassandra cluster is composed of several nodes forming a
logical topology of a ring. Each server node being assigned an arbitrary
(random) position in the ring. The dataset is distributed among these nodes by
deriving a position in the ring from a hash of each data item's key, and
assigning that item to the first node with a greater position.  In this way the
data items are ``wound around'' the ring (multiple times) and one node becomes
responsible for multiple ranges of keys.
We refer to each such range as a \emph{bucket} and use this level of
granularity for synchronization. That is, two transactions conflict if they try
to access the same bucket.
We show this model in \rfig{fig:arch}.

The original inbox application data model consists of two tables distributed in a
Cassandra cluster, one for term search and one for interaction search. Each
table row is a separate data item containing data from
all of the table's columns in that row (as the object's fields). 
The users of the application all have unique identifiers ({\tt UserID}), as do
all messages in the system ({\tt MsgID}). 
Note that the data resolution changes between the original application and the
benchmark, since bucket-level granularity gives us a greater level of control
over contention.

The term search table ({\tt TermTable}) is used to find messages in a user's
inbox by keywords. The logical structure of the table consists of one column
containing a {\tt UserID} and one column for each possible keyword. Each keyword
column contains lists of message identifiers ({\tt MsgID}s) of messages that
include this keyword. More formally:
$${\tt TermTable}: {\tt UserID} \mapsto {\tt Keyword} \times ({\tt MsgID}~{\tt list})$$
This schema assumes that the number of users is orders of magnitude greater
than the size of one user's inbox. However, this is problematic for the
purposes of a benchmark, since it makes contention both lower and more
difficult to adjust. Hence, we decided to modify this table to increase the
number of keys by moving the keyword data from separate columns into the key.
In effect, the number of objects per user is increased and contention is no
longer strictly dependent on the number of users but can be controlled by
adjusting the population of keywords. Hence our schema looks as follows:
$${\tt TermTable}: {\tt UserID} \times {\tt Keyword} \mapsto ({\tt MsgID}~{\tt list})$$
Note that, since the benchmark only stores hashes of keys, the hashed keyword
information has to be kept elsewhere in addition to the key column, but we omit
this detail for clarity.

The interaction search table ({\tt InterTable}) stores conversations
between users. The table's key is the identifier of the user sending the
message. The table also contains one column per receiving user which holds the
IDs of the messages exchanged between the two users.
Thus:
$${\tt InterTable}: {\tt UserID} \mapsto {\tt UserID} \times ({\tt MsgID}~{\tt list})$$
This schema encounters the same problems as {\tt TermTable}, so we perform a
similar modification in order to increase the key domain: we use an ordered pair
indicating the sender and the receiver of the message as the key, rather than
just the sender. Thus, in the benchmark the table is defined as:
$${\tt InterTable}: {\tt UserID} \times {\tt UserID} \mapsto ({\tt MsgID}~{\tt list})$$

We extend the data model by adding additional tables which can be used to
introduce new functionality into the application, and in this way increase the
benchmark's depth---widen the range of transactions that can be performed.
The message table ({\tt MessageTable}) stores the contents of the messages used
in the system (note that {\tt TermTable} and {\tt InterTable} both operate on
message identifiers alone). The table's key is the identifier of the recipient
of the message, and the value is the message, which can be expressed as five
columns: the identifier of the message, the user identifiers of the sender and
the recipient, the text of the message, and the timestamp of when it was sent.
(The timestamps are used to sort messages, allowing transactions to return a
group of most recent ones).
In effect, we have the following definition:
$$\begin{array}{l}
     {\tt MessageTable}: {\tt UserID} \mapsto
     ({\tt MsgID} \times {\tt UserID} \times {\tt UserID} 
                 \times {\tt Text} \times {\tt Time})~{\tt list}
\end{array}$$

Finally, we employ a sequential number table ({\tt SeqNoTable}) which is used
to generate new message identifiers for a given user's inbox and to specify a
cut off point for deleting old messages. Thus, the table's key is the
identifier of an owner of an inbox and the data columns include the sequence
number ({\tt SequenceNo}) of the most recently created message in the inbox and the
sequence number of the last deleted message there. This is defined as:
$${\tt SeqNoTable}: {\tt UserID} \mapsto {\tt SequenceNo} \times {\tt SequenceNo}$$

\subsection{Tasks and Transactions}
\label{sec:transactions}

Clients in the benchmark can run 8 types of high-level tasks on the distributed
database. Each of these tasks performs some non-transactional operations
(primarily locating nodes in the ring from hashes and performing local
processing) and employs one or more atomic transactions. The first two tasks:
term search and interaction search are a part of the original example in
\cite{LM10}, and the remainder are domain-appropriate additions that ensure the
configurability of the benchmark. We describe the tasks below and provide detail
about the transactions they use. 
All transactions are further detailed in \rfig{fig:transactions}.
We also provide a summary in \rfig{fig:summary}.

\begin{figure}[t]
\begin{minipage}[t]{0.49\textwidth}
\begin{lstlisting}
atomic getAssociation(userID1, userID2) {
  result[MSG_ID].addAll(InterTable[(userID1, userID2));
  result[MSG_ID].addAll(InterTable[(userID2, userID1));
  result[SEQ].add(SeqNoTable[userID1]);
  result[SEQ].add(SeqNoTable[userID2]);
}

atomic getByKeyword(myUserID, myKeywords) {
  for (myKeyword : myKeywords) 
    for ((userID, keyword) : TermTable)
     if (userID == myUserId and myKeyword == keyword) {
       msgIDs = TermTable[(userId, keyword)];
       result.addAll(msgIDs);
     }
}

atomic getConversation(senderID, recipientID) {
  for ((sendID, recpID) : InterTable)
    if (senderID == sendID and recpID == recipientID)
      result.addAll(InterTable[(senderID, recipientID));
}

atomic getMessages(msgIDs) {
  for (msgID : msgIDs)
    for (recipientID : MessageTable)
      if (recipientID == msgID.recipientID) {
        messages = MessageTable[recipientID];
        for (msg : messages)
          if (msgID == msg.msgID)
            result.add(msg);
      }  
}

atomic indexMessages(queries) {
  for ((userID, keywords) : queries)
    for (keyword : keywords) {
      msgIDs = TermTable[(userID, keyword)];
      for (msgID : msgIDs) {
        messages = MessageTable[msgID.recipientID];
        for (msg : messages)
          if (msgID == msg.msgID)
             result[MSG].add(msg);
      }
     }
   result[MSG].sortByTimestamp();
   for (userID : queries) 
     result[SEQ].add(SeqNoTable[userID]);
}
\end{lstlisting}
\end{minipage}
\begin{minipage}[t]{0.49\textwidth}
\begin{lstlisting}[firstnumber=49]
atomic resetCutoff(userID) {
  (currentSeq, deletedSeq) = SeqNoTable[userID];
  result.add((currentSeq, deletedSeq))
  SeqNoTable[userID] = (currentSeq, currentSeq);
}

atomic sendMsg(senderID, recipientID, content, keywords) {
  SeqNoTable[recipientID] += 1;
  (currentSeq, deletedSeq) = SeqNoTable[recipientID];
  msgID = new MsgID(recipientID, currentSeq);
  timestamp = currentTimestamp();
  msg = new Message(msgID, senderID, recipientID, 
                    content, timestamp);
  MessageTable[recipientID].add(msg);
  InterTable[(senderID, recipientID)].add(msgID);
  InterTable[(recipientID, senderID)].add(msgID);
  for (keyword : keywords) 
    TermTable[(recipientID, keyword)].add(msgID);
}

atomic removeMessages(messages) {
  for (msg : messages) {
    for (keyword : msg.keywords)
      TermTable[(msg.userID, keyword)].remove(msg.msgID);
    InterTable[(msg.senderID, msg.recipientID)].remove(msg.msgID);
    InterTable[(msg.recipientID, msg.senderID)].remove(msg.msgID);
    MessageTable[(msg.recipientID)].remove(msg.msgID);
  }
}

atomic importMessages(messages) {
  for (msg : messages) {
    (currentSeq, deletedSeq) = SeqNoTable[msg.senderID];
    if (msg.sequenceNo < deletedSeq)
      continue
    if (MessageTable[msg.recipientID].exists())
      continue
    MessageTable[msg.recipientID].add(msg);
    InterTable[(msg.senderID, msg.recipientID)].add(msgID);
    InterTable[(msg.recipientID, msg.senderID)].add(msgID);
    for (keyword : msg.keywords) 
      TermTable[(msg.recipientID, keyword)].add(msg.msgID);
  }
}
\end{lstlisting}
\end{minipage}
\caption{\label{fig:transactions}Benchmark transactions pseudocode.}
\end{figure}
\paragraph{Term Search} %

A user (identified by a specific {\tt UserID}) specifies a set of keywords. The
user's inbox is then searched to find all messages which contain one of the
keywords by using transaction {\tt getByKeyword}. 
This transaction traverses {\tt TermTable} and returns all message identifiers
whose key both matches the user's ID and at least one of the specified
keywords. This section must be atomic to return a consistent state of the
inbox.
Finally, the contents of the matching messages are retrieved with transaction
{\tt getMessages}, which retrieves the contents of the messages from {\tt
MessageTable} on the basis of the list of message identifier. Again, this is
done atomically, in order to return a consistent snapshot of the inbox.
Overall, the term search task reads from two tables: {\tt TermTable} and {\tt
MessageTable} and we estimate it to be medium in length. The length of the task
will depend on the size of the keyword domain.

\paragraph{Interaction Search}

A user specifies a {\tt UserID} of another user in order to find all conversations
between the two of them. %
In order to do this, the task runs transaction {\tt getConversation}, which gets
all message identifiers from {\tt InterTable} whose keys fit the user
identifiers of both users in question. Then, the contents of the messages are
retrieved using {\tt getMessages}.
Thus, interactions search reads from two tables: {\tt InterTable} and {\tt
MessageTable} and is of small length, depending on the number of users in the
system and the number of messages in both the users' inboxes.

\paragraph{Send Unicast}

A user, the sender, sends a message consisting of a text content to another
user, the recipient. This consists of executing transaction {\tt sendMsg} which
is preceded by extracting the set of keywords from the contents of the message.
The transaction then increments and retrieves the next sequence number from
{\tt SeqNoTable} and uses it to create a new {\tt MsgID}. Then, a message
object is created and written into {\tt MessageTable}. Subsequently, the
identifier of the message is inserted into {\tt InterTable} twice: once to
indicate the message sent out from the senders inbox, and once to indicate the
message received in the inbox of the recipient. Finally, the transaction
inserts an entry into {\tt TermTable} for each keyword extracted from the
contents of the message. All these actions must be performed atomically, in
order to prevent another transaction from interfering and causing an
inconsistent state. E.g., if a message were removed by another transaction from
the {\tt MessageTable} while this task was yet to add all the keywords to {\tt
TermTable}, the contents of {\tt TermTable} would point to messages which no
longer existed.
Overall, this task updates {\tt SeqNoTable} and writes to all the other tables.
On the whole, this is a short read/write task with a relatively limited
read/write set (R/W set).

\paragraph{Send Multicast}

A user sends a single message to several recipients. This task is an extension
of the send unicast task (defined above), where the {\tt sendMsg} transaction
is executed in series, but the entire series is executed atomically.
This task is a medium- to large-sized write task that touches the same tables as the send
unicast task, but it strongly depends on the number of recipients.

\paragraph{Batch Import}

The system is given a set of complete messages to store in the database. This
represents a use case where the database is replicated and two of the replicas
synchronize, by one sending a state update to the other. Another pertinent
scenario is one where the database tries to recover from a crash.
The task involves extracting keywords from each message's contents and
executing transaction {\tt importMessages}. The transaction filters the set of
imported messages against {\tt SeqNoTable}  to remove all those that have a
sequence number lower or equal to the sequence number of the last deleted
message. If a message has a higher sequence number than the current highest
sequence number in {\tt SeqNoTable}, then the database is also updated. Then,
the transaction filters out those messages which are already in {\tt
MessageTable}. All the remaining messages are added to {\tt MessageTable}, {\tt
InterTable}, and {\tt TermTable} by analogy to {\tt sendMsg}.
The task is a long write task (depending on the number of imported
messages) which updates {\tt SeqNoTable} and {MessageTable}, and writes to the
two remaining tables.

\paragraph{Clear Inbox}

A user requests that all messages from her inbox be removed. This task
sequentially executes three separate transactions. 
First, the task runs transaction {\tt resetCutoff} which retrieves the sequence
numbers of the most recent message in the user's inbox as well as the sequence
number of the most recent deleted message from {\tt SeqNoTable}. In addition,
the transaction sets the deleted sequence number to the current sequence
number, signifying that all messages in the inbox are now below the cutoff for
deletion.
Next, the task uses transaction {\tt getMessages} to retrieve all the existing
messages in the inbox.
Afterward, the task executes transaction {\tt removeMessages} to remove all
the messages from the three tables in the database that hold message
information. Specifically, for each keyword of each message to be removed, the
transaction removes each message's identifier from {\tt TermTable}.  Then, on
the basis of the recipient and sender of each message, the transaction removes
both occurrences of its message identifier from {\tt InterTable}. Finally, the
transaction searches each message in {\tt MessageTable} by their recipient and
remove them from there.
The task is a medium-sized read/write tasks (although the execution can be shorter
depending on the number of messages in the inbox) and touches all tables.  The
task is not executed atomically as a whole because any potential
inconsistencies stemming from the lack of atomicity between transactions can be
fully and easily resolved by the application. Thus this decomposition makes for
a more realistic workload. 

\paragraph{Association Level}

Given two users, the system checks what is the level of interaction between
them, by counting the number of exchanged messages and normalizing the number
against the number of messages in the inbox. This involves running transaction
{\tt getAssociation}, which retrieves the identifiers of messages involved in 
conversations between both users (from {\tt InterTable}). It also retrieves the
sequence numbers of the most recent existing and the most recent deleted
message in both users' inboxes (from {\tt SeqNoTable}).
The results of these executions are then processed non-transactionally by the
task. This produces a short read-only task operating on a single table---{\tt
InterTable}. We add this task to have a finer control over contention, since
the existing short read-only transactions in interaction search and term search
are always followed by a medium-sized read-only transaction.

\paragraph{Indexing}

Creates a cache for the most common keyword searches by the most active users.
Given a list of users and a list of keyword search queries per user the task
runs transaction {\tt indexMessages}. For each user and for every keyword, the
transaction searches through {\tt TermTable} and collects message identifiers
of all pertinent messages. Then, unique message identifiers are extracted from
the list. The results are then sorted and cropped to a prescribed length. Then,
the transaction retrieves the body of the three example messages for each user from {\tt
MessageTable}. Finally, the transaction retrieves sequence number information from {\tt
SeqNoTable} for each of the investigated users.
Indexing is a work-intensive read-only task that contains a long transaction
with a large readset.

\begin{figure}[t]
{\footnotesize
\begin{center}

\begin{tabular}{|l|l|l|l|}
\hline
Task               & R/W set    & Length       & Used transactions \\ %
\hline
\hline
association level  & read-only  & short        & {\tt getAssociation}    \\ %
term search        & read-only  & medium       & {\tt getByKeyword}, 
                                                 {\tt getMessages}       \\ %
interaction search & read-only  & short        & {\tt getConversation},
                                                 {\tt getMessages}       \\ %
indexing           & read-only  & long         & {\tt indexMessages}     \\
send unicast       & read/write & short        & {\tt sendMsg}           \\ %
send multicast     & read/write & medium/long  & {\tt sendMsg}           \\ %
batch import       & read/write & long         & {\tt importMessages}    \\ %
clear inbox        & read/write & medium         & {\tt resetCutoff},      
                                                 {\tt getMessages},      
                                                 {\tt removeMessages}    \\ %
\hline 
\hline
Transaction           & R/W set  & Length      & Touched tables       \\ 
\hline
\hline
{\tt getByKeyword}    & read-only  & short/medium & {\tt TermTable} \\
{\tt getConversation} & read-only  & short       & {\tt InterTable} \\
{\tt getAssociation}  & read-only  & short       & {\tt InterTable} \\
{\tt getMessages}     & read-only  & medium      & {\tt MessageTable} \\
{\tt indexMessages}   & read-only  & long        & {\tt TermTable}, 
                                                   {\tt SeqNoTable} \\
{\tt resetCutoff}     & read/write & short       & {\tt SeqNoTable} \\
{\tt sendMsg}         & read/write & medium      & {\tt TermTable}, 
                                                   {\tt InterTable}, 
                                                   {\tt MessageTable}, 
                                                   {\tt SeqNoTable} \\
{\tt importMessages}  & read/write & long        & {\tt TermTable}, 
                                                   {\tt InterTable}, 
                                                   {\tt MessageTable},    
                                                   {\tt SeqNoTable} \\
{\tt removeMessages}  & read/write & medium      & {\tt TermTable}, 
                                                   {\tt InterTable}, 
                                                   {\tt MessageTable} \\
\hline
\end{tabular}
\end{center}
\caption{\label{fig:summary}Overview of tasks and transactions in \benchmark.}
}
\end{figure}

\subsection{Metrics}
\label{sec:metrics}

In recognition of the fact that complex systems require comprehensive metrics,
the benchmark implementation includes several metrics useful for monitoring
distributed TM performance.

Classically, we provide the commonly employed \emph{throughput} metric, defined
as the number of transactions executed per second by the system. 
We also measure \emph{latency} (or \emph{mean flow time}), which is an
alternative performance goal to throughput. A single transaction's latency
(\emph{flow time}) is the time between a transaction commences and finally
completes. This time includes any time spent re-executing the transaction due
to forced aborts. The system's metric is the mean of all transactions' latency.
We measure latency on the system level as well as per transaction.
Furthermore, we allow measuring transactions' \emph{abort rate}, which
indicates how often transactions are forced to re-execute.
These metrics allow to measure to what extent a TM's overall performance goals
were reached. They also constitute the most basic tool for comparing different
TMs. 

In addition, we provide more diagnostic metrics that allow to reason about the
workload itself. 
First of all, we measure \emph{retry rate}, \emph{total execution time of all transactions},
\emph{total execution time of all retries}, \emph{sum of all startup times for all transactions} i.e.
sum of all periods of time between the start of a transaction and the start of the first retry
of this transaction, as well as \emph{total number of operations on all the buckets}.
Finally, we supplement these measures with metrics showing the \emph{total
execution time} of the system, \emph{total parallel execution time} i.e. time between the start
of the first client thread and the end of the last client thread, and \emph{transactional execution
ratio}, the percentage of time the system spent within transactions.

\subsection{Parameters}
\label{sec:parameters}

The benchmark includes a number of parameters that can be used to evaluate a TM
in a range of workload types. 
The most important aspect controlled by the parameters is contention, which can
be controlled by adjusting the number of shared objects in relation to the
number of transactions that use these objects on average, and to the number of
objects used by each transaction. Larger R/W sets or more clients in relation
to the same number of objects means increased contention.

We allow fine control over the domain of shared objects by changing the
configuration of the  system: the number of network nodes, and the number of
remote objects (buckets) per table. We also allow changing the domain of words
allowed in messages, which increases the number of keys in {\tt TermTable}.
The number of transactions is controlled by adjusting the number of
simultaneous clients in the system, as well as the number of tasks executed by
each of them. 
Finally, the R/W sets of transactions can be controlled directly by adjusting
the cap on the number of keywords allowed in a query ({\tt getByKeyword}, {\tt
indexMessages}), setting the minimum and maximum number of messages for {\tt
importMessages}.

The benchmarks also allow to control the composition of the workload, by
specifying what percentage of which types of transactions will be executed.
This allows to test various features of TM memories that depend on e.g., a high
ratio of read operation or read-only transactions.
In addition, since moving data around a network can have a significant cost, we
allow adjusting the size of data by setting the maximum length of the content
of messages in {\tt MessageTable}. Hence caching and storage strategies used by
various distributed TMs can be evaluated in various environments.
Lastly, we introduced a configurable delay in milliseconds that is applied to
every operation on every bucket in the system. It can be used to mimic higher
network latency or slower nodes.

%% file: evaluation.tex
\section{Evaluation}
\label{sec:evaluation}

In this section we present an example of empirical evaluation of several
distributed concurrency control mechanisms, including two distributed TMs
representative of two divergent approacches to synchronization, using
\benchmark{}. 
We must stress, however, that the purpose of the paper is not to provide a
comprehensive evaluation of the TM systems in question (both of which were
evaluated in their repsectve papers \cite{SW15-ijpp,TRP13}). 
Instead, the main objective of this evaluation is to provide a set of reference
parameters and workloads and demonstarte how particular settings impact the benchmark's
workload and the performance of disparate concurrency control schemes. In
effect, this shows that \benchmark{}' parameters are impactful and that the
benchmark can be used to mimic complex, real-life distributed systems.
In addition, we showcase how an evaluation can be used to draw meaningful
diagnostic conclusions and provide a reference benchmark for future research in
distributed TM.

\subsection{Frameworks}

For the purpose of comparison we use four distributed concurrency control
frameworks, including two distributed TM implementations that represent two
main approaches to transactional synchronization, and two typical distributed
locking schemes.

The first distributed TM we evaluate is HyFlow2 \cite{TRP13}, a state-of-the-art
optimistic data-flow distributed TM implemented in Scala. The optimistic
approach to concurrency control means that when two transactions conflict on a
shared object, one of them is aborted and re-executed. The data-flow model
means that whenever a client accesses a shared object, the object is moved to
the client's node for the duration of the access (but there is always exactly
one copy of each object in the system). HyFlow2 implements the Transaction
Forwarding Algorithm (TFA) 
\cite{SR12} 
to handle synchronization (with the
guarantee of opacity \cite{GK10}) and uses the Akka library for networking. For
the evaluation we configure HyFlow2 to use standard Java serialization.

The second distributed TM is Atomic RMI \cite{SW15-ijpp}, a pessimistic
control-flow distributed TM implemented in Java, on top of Java RMI. The
pessimistic approach means transactions defer operations to avoid conflicts
altogether and thus preventing transactions from aborting. The control flow
model means that shared objects are immobile and execute any code related to
accesses on the nodes they are on. Atomic RMI uses the Supremum Versioning
Algorithm (SVA) 
\cite{Woj07} 
for concurrency control, which guarantees last-use
opacity \cite{SW14-disc}.

The two locking schemes used for the evaluation are fine grained locks (FGL)
and a global lock (GLock). FGL simulate an expert implementation of a
distributed locking scheme using mutual exclusion locks with one lock per
shared object. Locks in FGL are always acquired and released according to
two-phase locking (2PL) and used according to a predefined global locking order
to prevent deadlocks. Locks are released in FGL after the last access to a
particular object within a given transaction. Such an implementation of 2PL is
trivially last-use opaque.
GLock is a locking scheme where the system contains a single lock on one of the
nodes that must be acquired at the beginning of each transaction and held
throughout each transaction's duration. This means that only one transaction
can be executed at a time in the entire system. Hence, this is an anchor
implementation, roughly equivalent to using a sequential baseline execution in
non-distributed TM evaluations. GLock is trivialy opaque.
Both locking schemes used for the evaluation are built on top of Java RMI and
use custom lock implementations.

\subsection{Testing Environment}

We perform our evaluation using a 16-node cluster connected by a 1Gb network.
Each node is equipped with two quad-core Intel Xeon L3260 processors at 2.83
GHz with 4 GB of RAM each and runs a OpenSUSE 13.1 (kernel 3.11.10, x86\_64
architecture). We use the 64-bit Java HotSpot(TM)
JVM version 1.8 (build 1.8.0\_25-b17).

\subsection{Parameter Settings}

\input{ScenariosTable}

As the first part of our evaluation we showcase the benchmark's configurability
by displaying the behavior of the four frameworks within diverse workloads
generated by manipulating the parameters of \benchmark{}.
The starting point for the reconfiguration is the \emph{standard} workload,
which has an 80--20\% read-to-write-task ratio (the probabilities of executing a
specific task are shown in \rfig{fig:tasks}) typically used for TM evaluation. 
The basic configuration of the standard scenario contains 1024 buckets deployed
on 16 nodes, with 200 concurrent clients executing 3 consecutive tasks each.
The message length is set to 8 words and the operation delay is set to 3ms.
We treat these parameters as our reference for comparing TM implementations.
The parameters were selected for the standard benchmark on the basis of
experimentation to simulate a well-behaved distributed TM system. 
We then show how the manipulation of these reference values impacts the
performance of the various concurrency control mechanisms.

\begin{figure}[t]
\subfloat[\label{fig:buckets}]
        {\includegraphics[width=\linewidth]{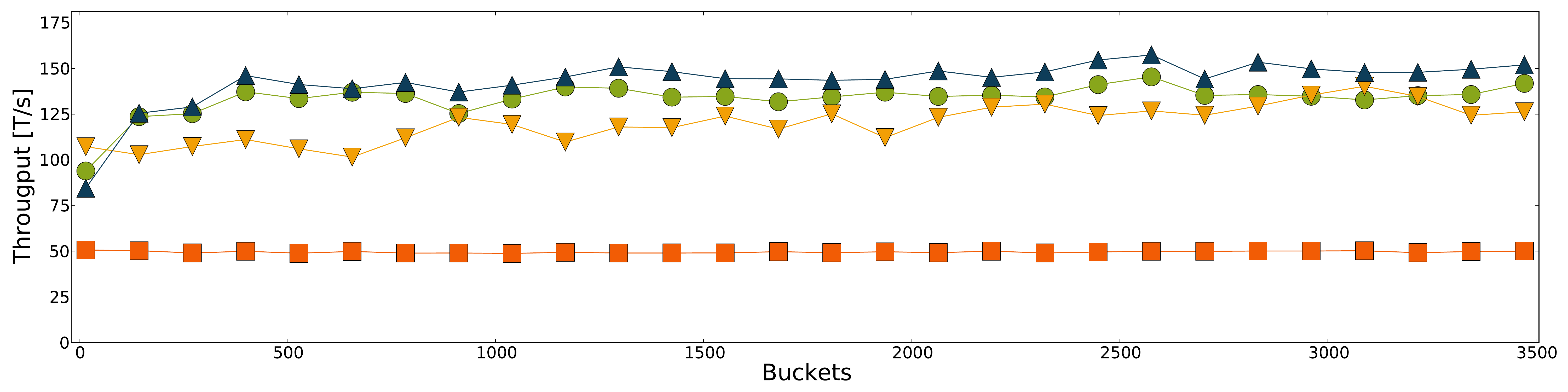}}

\subfloat[\label{fig:delay}]
        {\includegraphics[width=.33\linewidth]{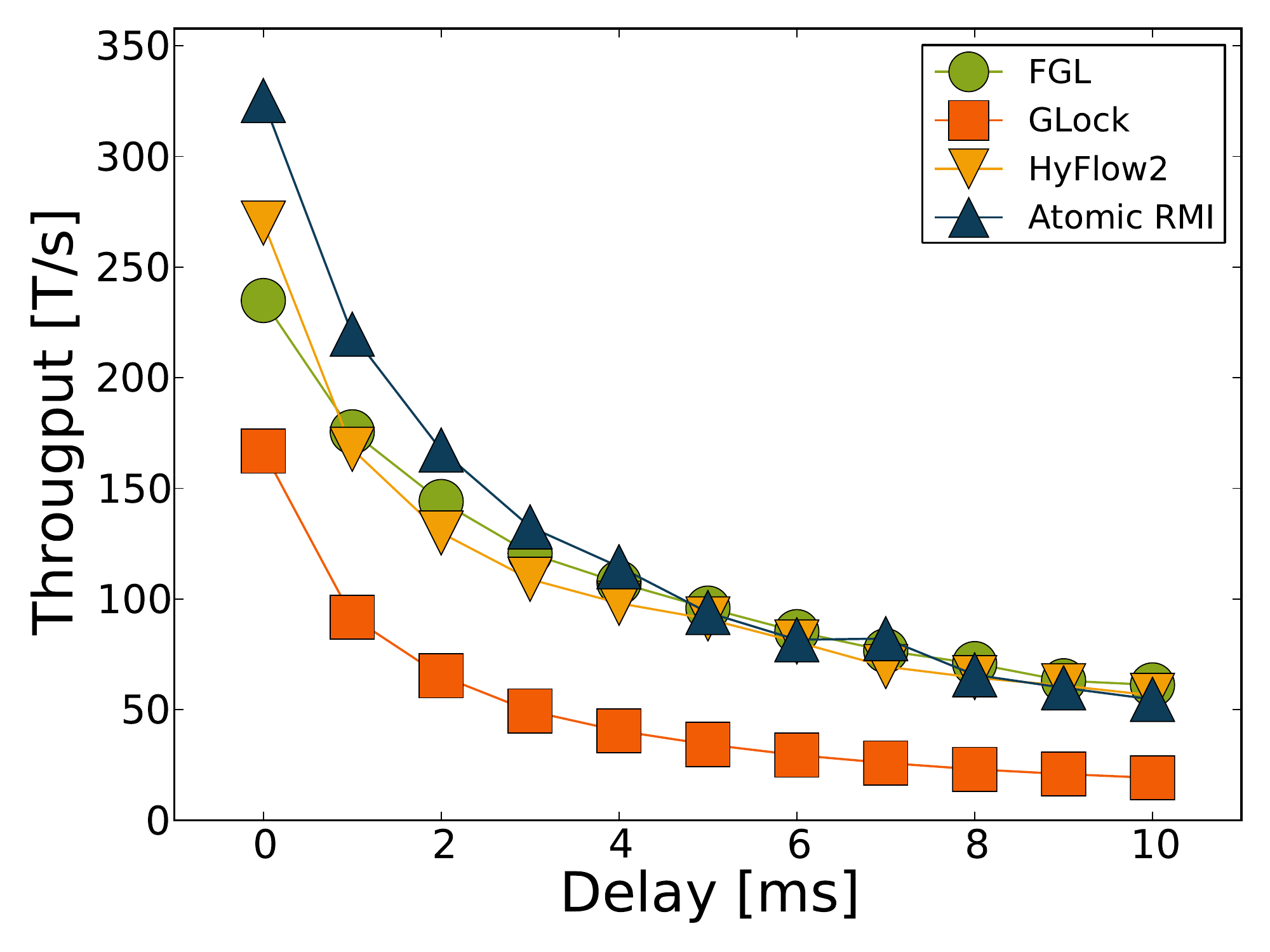}}
\subfloat[\label{fig:jobs}]
        {\includegraphics[width=.33\linewidth]{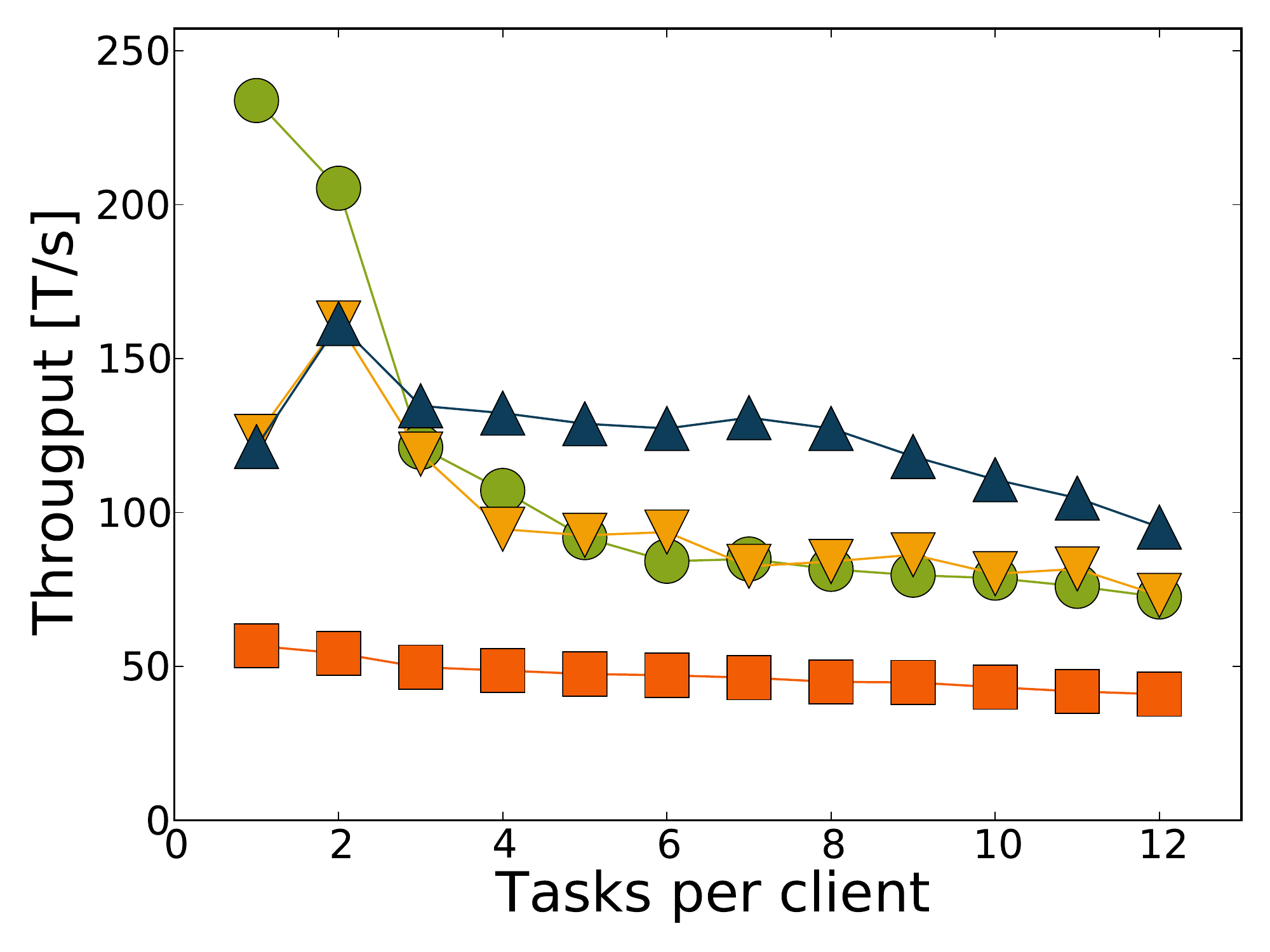}}
\subfloat[\label{fig:nodes}]
        {\includegraphics[width=.33\linewidth]{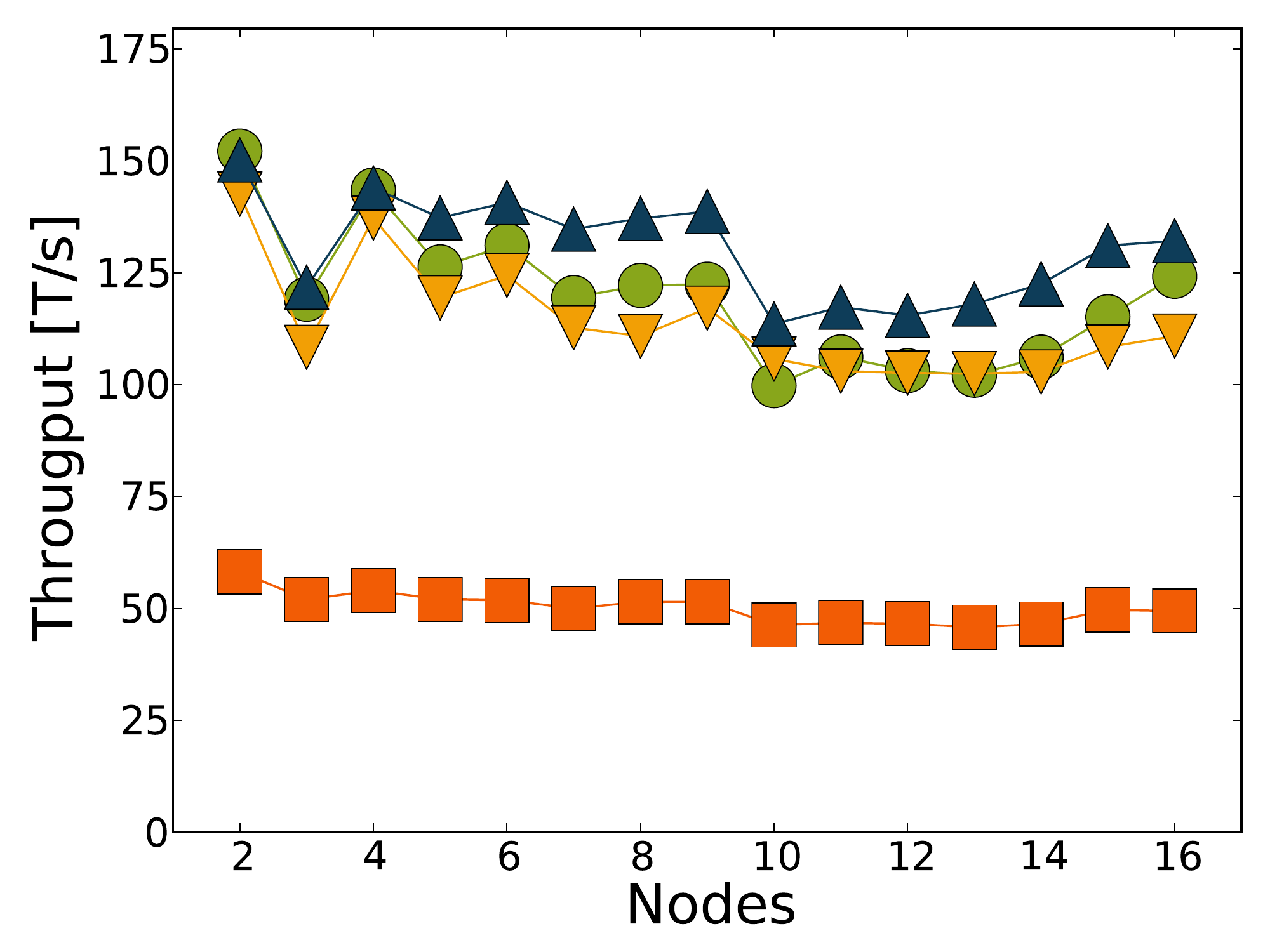}}
\caption{\label{fig:parameters}Impact of parameters on throughput.}
\end{figure}

\paragraph{Buckets} \rfig{fig:buckets} shows the impact of manipulation of the
number of buckets used within the system on throughput. As the number of
buckets is increased, transactions are less likely to access the same bucket,
so the system's contention decreases. A decrease in contention leads to an
increase in throughput for all evaluated frameworks, but the change has varying
impact. The evaluation shows that Atomic RMI and FGL both capitalize on higher 
contention in the range between 1 and 500 buckets, which is attributable to the
higher costs of conflict avoidance in those frameworks.  This is contrasted
with HyFlow2, whose throughput increases more gradually in the same range, as
contention lowers. When the number of buckets exceeds 1000, any further
increase does not result in an improvement in throughput, showing that at this
point contention is negligible and other factors, like operation overhead,
dictate performance.
Given these results, we use 1024 buckets as the reference value.

\paragraph{Delay} \rfig{fig:delay} shows the transactional throughput of the
four frameworks as the length of the operations executed by transactions
changes. Naturally, as the length of each operation increases, the throughput
of the system decreases (even for a sequential execution simulated by GLock).
Here, all frameworks behave similarly, showing a steady decrease in
performance, that converges around the 10ms mark.
The reference delay value we select is 3ms, at which point the throughput for
FGL, HyFlow2, and Atomic RMI is in the vicinity of 175 transactions per
second, with a slight divergence.

\paragraph{Tasks per client} \rfig{fig:jobs} shows the change in transactional
throughput as the number of consecutive tasks executed by each client increases.
The results clearly show anomalous behavior for low task numbers, that
significantly diverges from the results for higher task values. The divergence
stems from the heterogeneity of tasks in \benchmark{}, which means that for low
task values some clients will finish their tasks much faster than their
counterparts that execute longer tasks at the same time. If the number of
consecutive tasks is low, the clients with shorter tasks finish execution early
and leave the system with a diminished number of concurrent clients, thus
randomly, and unwarrantably decreasing the system's contention. When the number
of tasks per client is increased, the results show that performance steadies
for all frameworks. On the other hand, increasing the task count leads to a
    prohibitive total execution time for the entire benchmark (e.g. for FGL,
    the execution time for each data point is 35.2s for 3 consecutive tasks,
    43.1s for 4, and 96.4s for 10). Hence, we select 3 tasks, the least value
    where the performance is not anomalous, for the reference value.

\paragraph{Nodes} \rfig{fig:nodes} shows the reaction of the four frameworks to
changes in network size. As nodes are introduced into the distributed system,
the contention and transactional characteristics remain constant, which causes
little change in the throughput values for all frameworks. The number of nodes,
along with the number of buckets, decides how buckets are distributed in the
system. Given a constant number of buckets, the lower the number of nodes, the
greater the number of buckets that are hosted per node. This change itself
does not impact the behavior of the TM frameworks themselves, however.  What
the fluctuation in throughput reflects is the impact of the physical properties
of the testing environment itself in general, and the network topology as well
as communication costs in particular, which is consistently more pronounced for
certain configurations (i.e. for a system with 3 nodes or 10 nodes).
The reference value for node count selected for the benchmark is 16, since the
TM applications are meant to be used in larger rather than smaller systems, and
that is the largest configuration available to us.

\begin{figure}[t]
\centering
\subfloat[\label{fig:ml4}Throughput with message length at 4 words.]
        {\includegraphics[width=.4\linewidth]{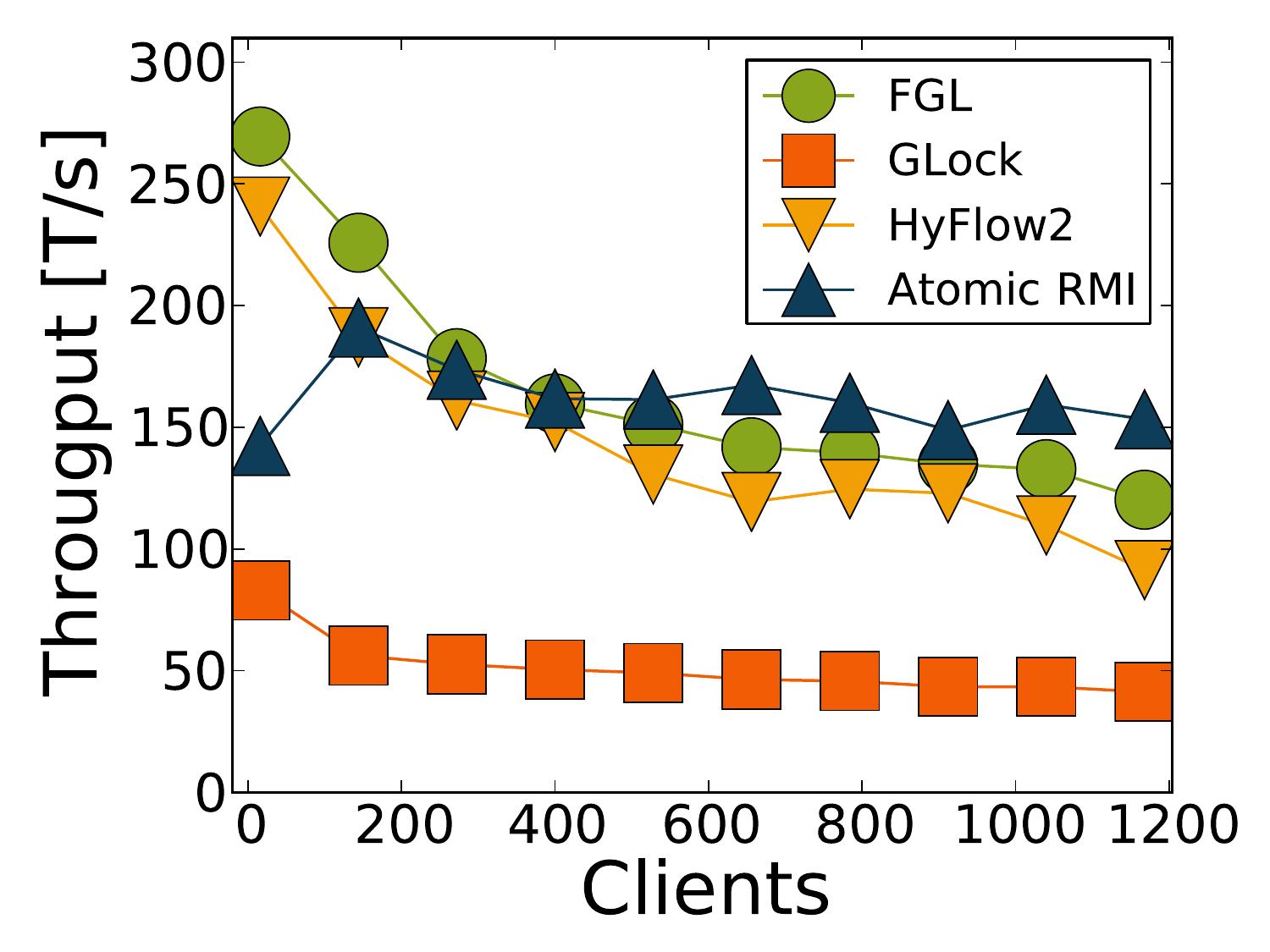}}
\hspace{1cm}
\subfloat[\label{fig:ml8}Throughput with message length at 8 words.]
        {\includegraphics[width=.4\linewidth]{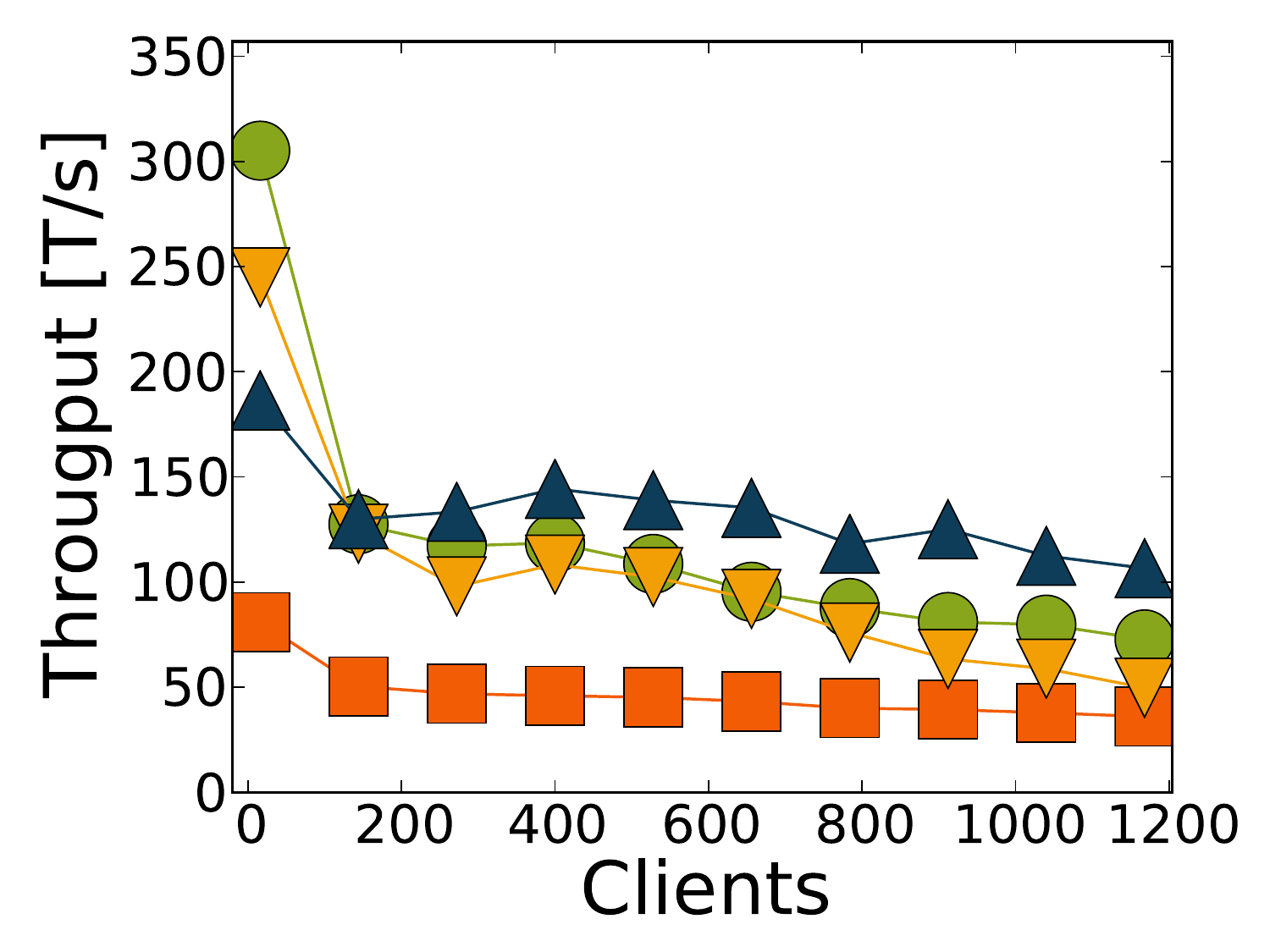}}
        
\subfloat[\label{fig:ml12}Throughput with message length at 12 words.]
        {\includegraphics[width=.4\linewidth]{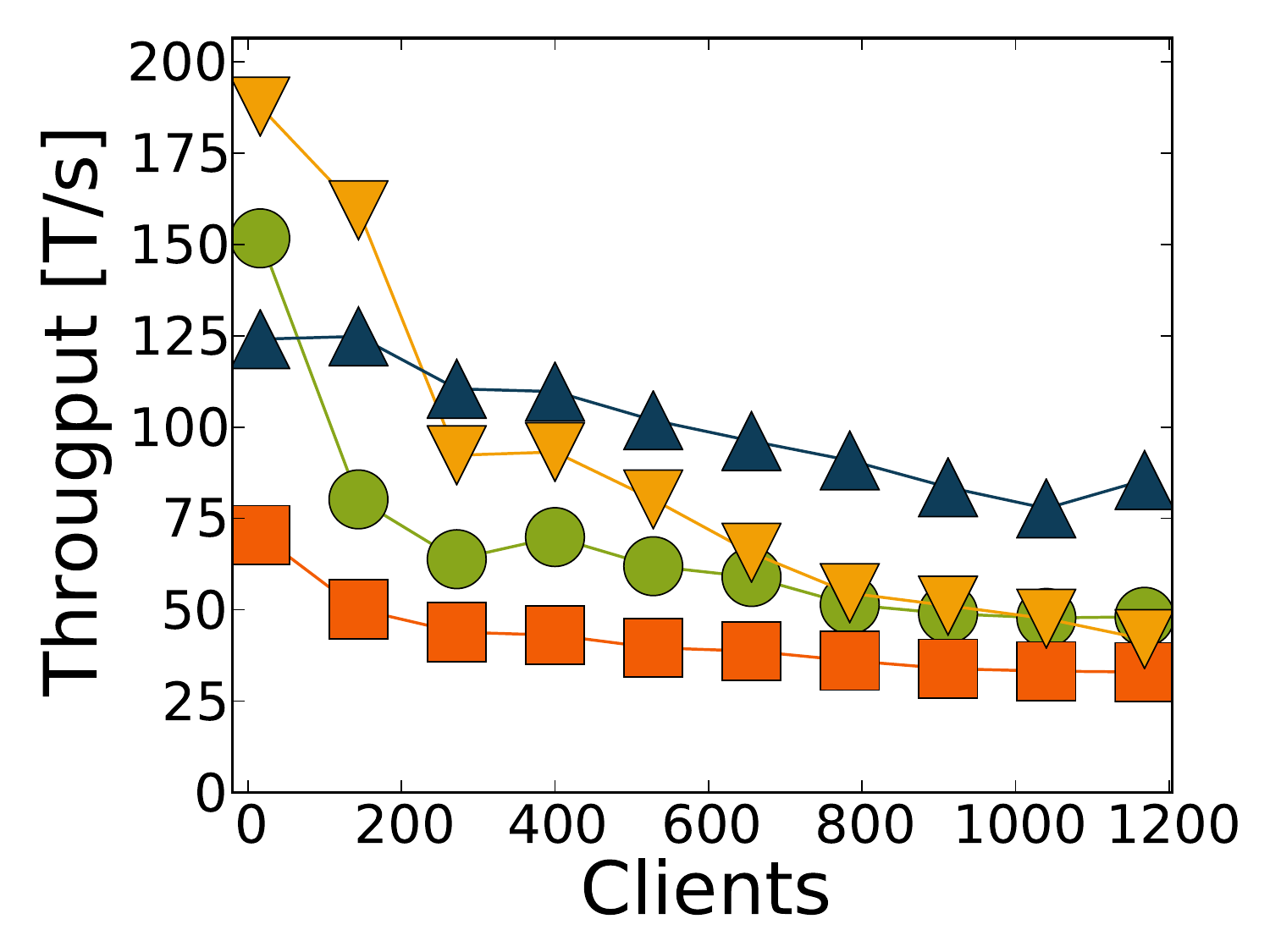}}
\hspace{1cm}
\subfloat[\label{fig:ml16}Throughput with message length at 16 words.]
        {\includegraphics[width=.4\linewidth]{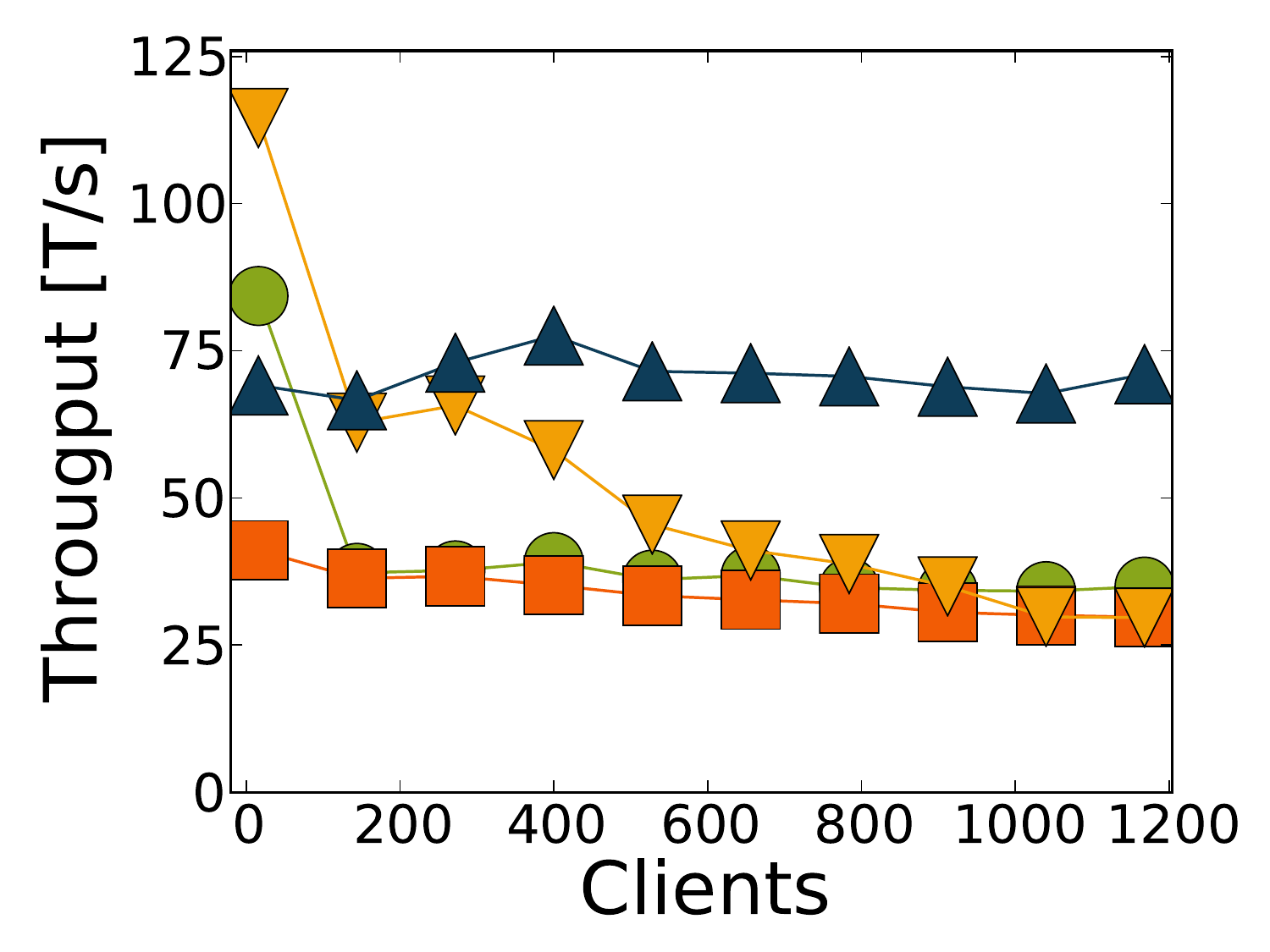}}
\caption{\label{fig:message-length}Impact of message length on throughput.}
\end{figure}

\paragraph{Message length} \rfig{fig:message-length} shows the impact of
setting a different message length for the contents of messages, as it affects
{\tt TermTable}. An increasing message length has a visible effect on all those
transactions that access {\tt TermTable}, since it increases the access sets of
transactions that read from that table (because parts of a message are
distributed among a greater number of buckets) and the length of write
operations on that table. Hence, increasing the message length impacts
contention, and thus decreases the transactional throughput of each framework.
The impact is most visible for FGL, where an increased message length causes a
significant drop in the presence of high client numbers.
The reference value for message length is set at 8 words.

\subsection{Scenarios}

\begin{figure}[t]
\subfloat[\label{fig:small-r}Small R scenario.]
        {\includegraphics[width=.33\linewidth]{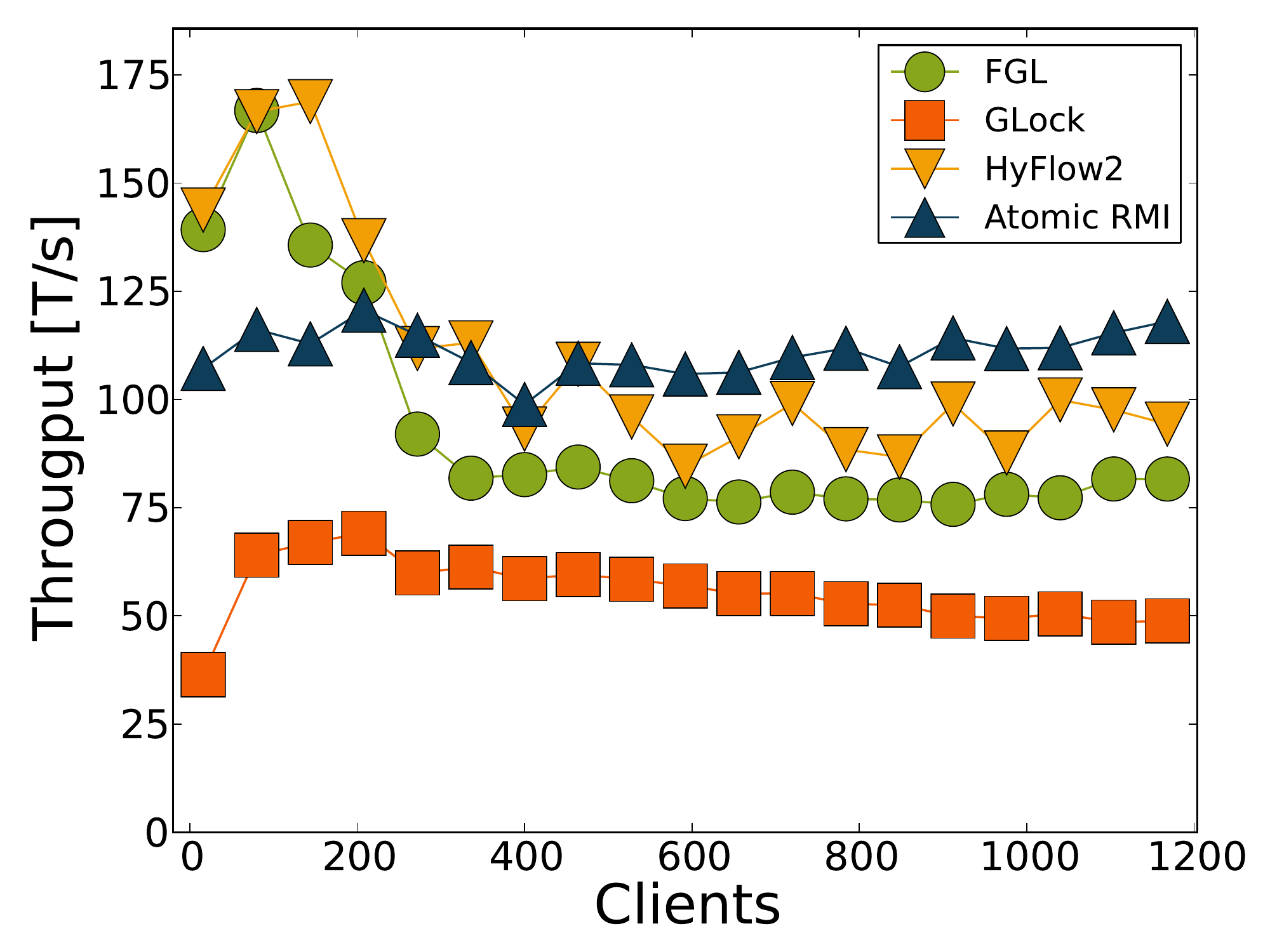}}
\subfloat[\label{fig:small-rw}Small R/W scenario.]
        {\includegraphics[width=.33\linewidth]{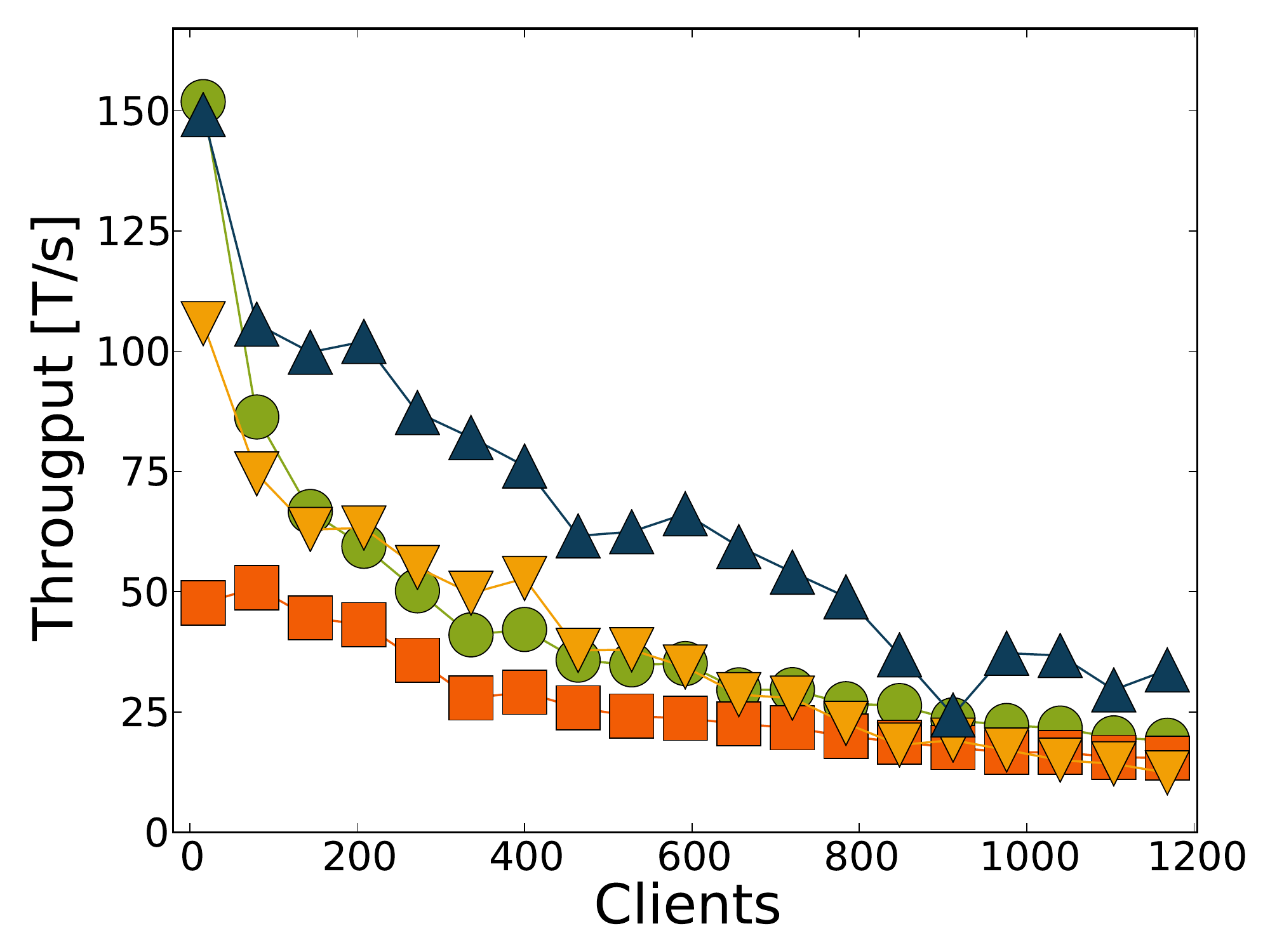}}
\subfloat[\label{fig:small-w}Small W scenario.]
        {\includegraphics[width=.33\linewidth]{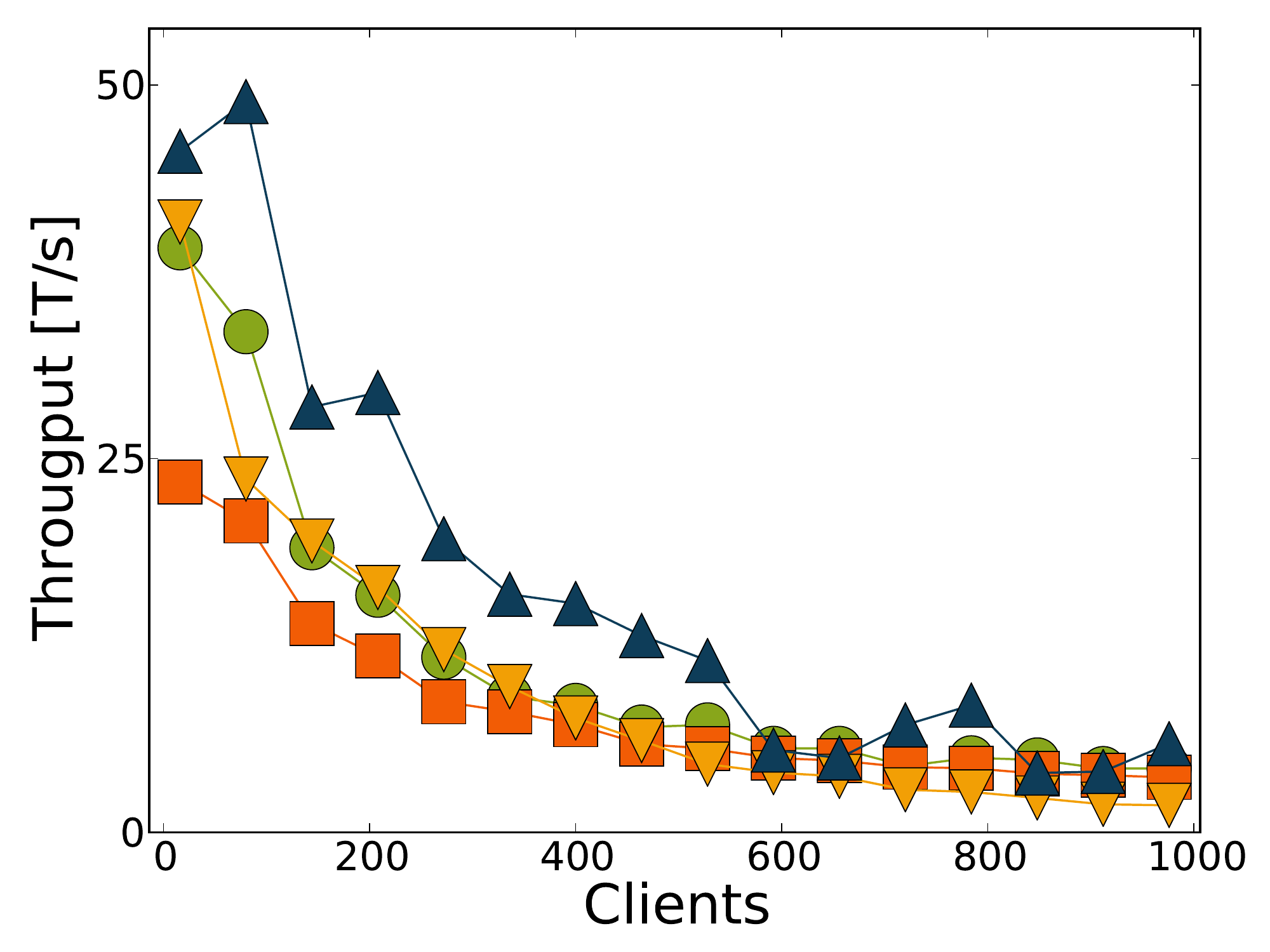}}

\subfloat[\label{fig:large-r}Large R scenario.]
        {\includegraphics[width=.33\linewidth]{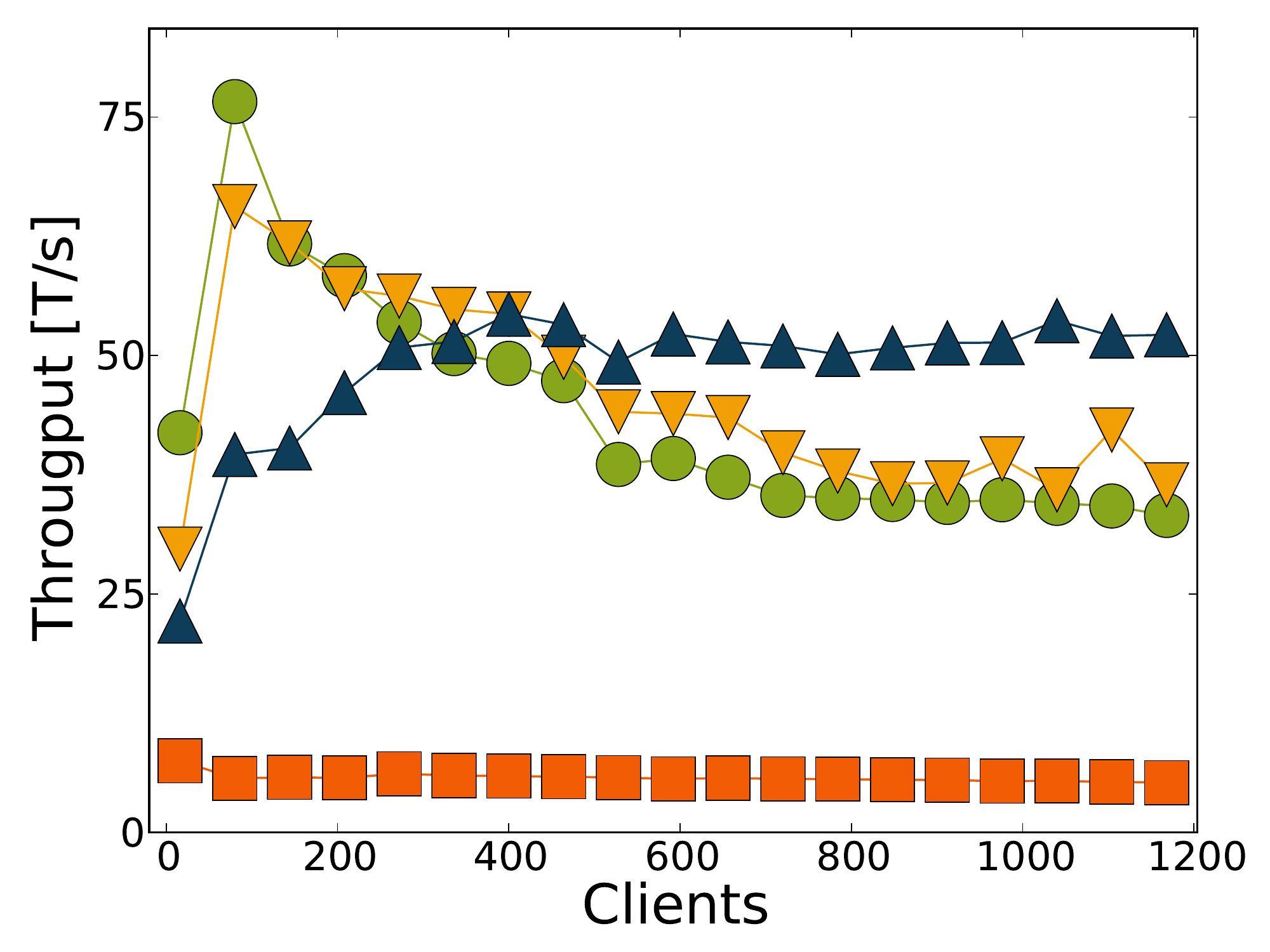}}
\subfloat[\label{fig:large-rw}Large R/W scenario.]
        {\includegraphics[width=.33\linewidth]{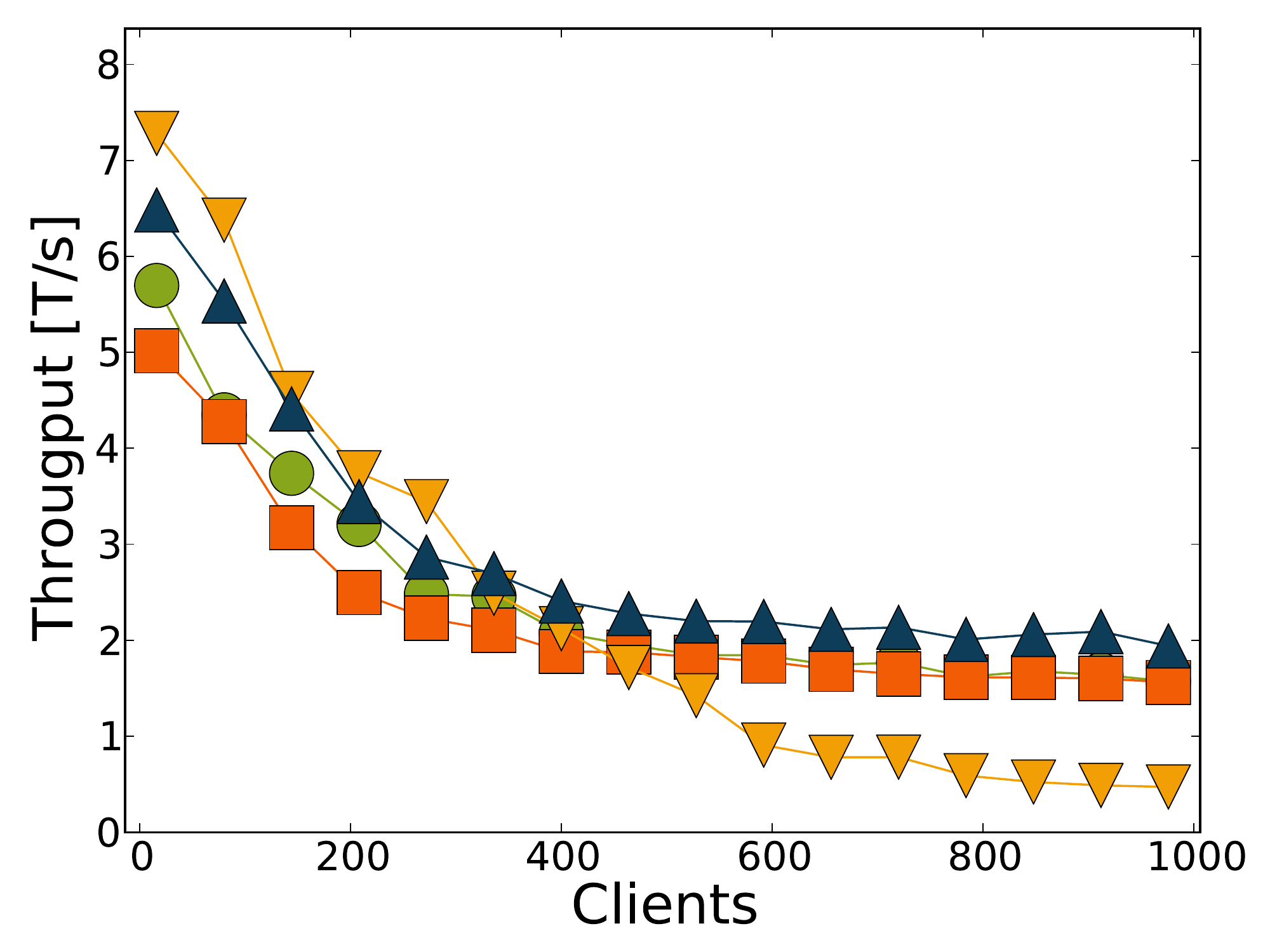}}
\subfloat[\label{fig:large-w}Large W scenario.]
        {\includegraphics[width=.33\linewidth]{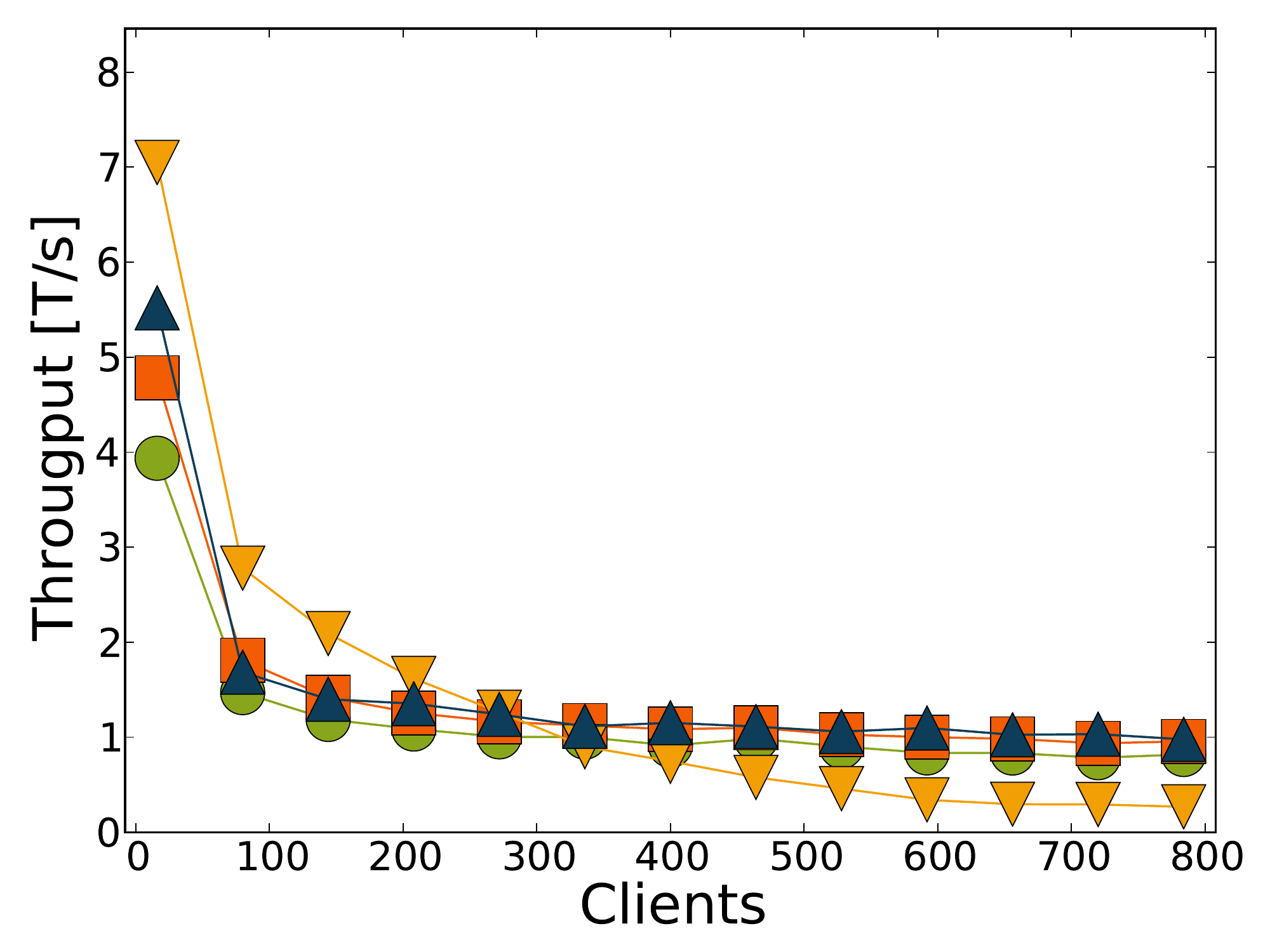}}
\caption{\label{fig:workloads}Workload configurations.}
\end{figure}

\rfig{fig:workloads} shows a comparison of the four benchmarks using different
workloads, wherein clients execute different sets of tasks. The composition of
particular scenarios in use is shown in \rfig{fig:tasks}, but generally,
scenarios marked \emph{small} consist primarily of short and medium tasks, with
a minimal number of long tasks, whereas scenarios marked \emph{large} consist
primarily of medium and long tasks, with sporadic short tasks also occurring.
Scenarios marked \emph{R} are read-dominated, consisting predominantly of tasks
executing read-only transactions, \emph{W} are write-dominated, consisting
mostly of tasks executing read/write transactions, and \emph{R/W} scenarios are
balanced.
Comparing the performance of the four frameworks using a varied set of
scenarios allows us to evaluate in a 
realistic environment and draw
conclusions from their performance.

In scenario \emph{Small R} (\rfig{fig:small-r}), we see that the increase in
contention driven by a growing number of clients impacts Atomic RMI least,
whereas the impact on FGL and HyFlow2 is significantly more visible.  HyFlow2
and FGL start off with an advantage in performance, with throughput as high as
250 transactions per second, until the number of clients reaches 200, when the
performance suddenly degrades and stabilizes between 75 and 100 transactions
per second for client counts larger than 300. In contrast, Atomic RMI retains a
constant throughput between 100 and 125 transactions per second regardless of
scale.
We see, however, that the aforementioned frameworks manage to scale in the
300--1200 client range with reasonable throughput, and retain an advantage of a
sequential execution of at least 25\% and up to 150\%.
The benchmark then shows that for specific sets of parameters and a particular
workload Atomic RMI maintains better scalability, whereas solutions like FGL
and HyFlow2 have a definite edge in lower contention environments.

In scenario \emph{Small R/W} (\rfig{fig:small-rw}) we see a general decrease in
throughput, as a large contingent of read/write operations are introduced into
the benchmark. These shape the transactional characteristic (e.g. for
all frameworks there is a gradual degradation in performance.
\rfig{fig:small-rw} and \rfig{fig:small-w}). Due to its operation-type agnostic
concurrency control method and an automated method of early release, Atomic RMI
handles the increase in write operations better than FGL and HyFlow until the
contention increases significantly with the introduction at 800 clients, at
which point all frameworks tend to converge and achieve a near-sequential
performance (up to 30 transactions per second for Atomic RMI vs 20 transaction
per second for GLock). Even though FGL is also pessimistic and has early
release, releasing objects early incurs additional network messages in that
scheme (since a lock has to be signaled), which, in turn, impacts performance
in a highly saturated network such as the one in this scenario.  On the other
hand, a large number of write operations and growing contention cause HyFlow2
to achieve a significant abort rate and a high retry rate (see below), causing
the transactions to often waste work and incur additional network traffic as
data-flow objects are supplied to transactions.
The results show that pessimistic TM may be favorable to optimistic TM in a
high contention write-dominated distributed system.

The \emph{Small W}  scenario (\rfig{fig:small-w}) is analogous to \emph{Small
R/W}, showing similar performance characteristics, but between 50\% and 75\%
decreased. It is plainly visible that for relatively low numbers of clients
(around 600) all frameworks tend to saturate and perform the same as or not
much better than a sequential evaluation.

\begin{figure}[t]
\begin{minipage}[b]{0.47\linewidth}\centering
\input{AbortRate_RetryRate}
\end{minipage}
\hspace{1cm}
\begin{minipage}[b]{0.47\linewidth}\centering
\includegraphics[width=\linewidth]{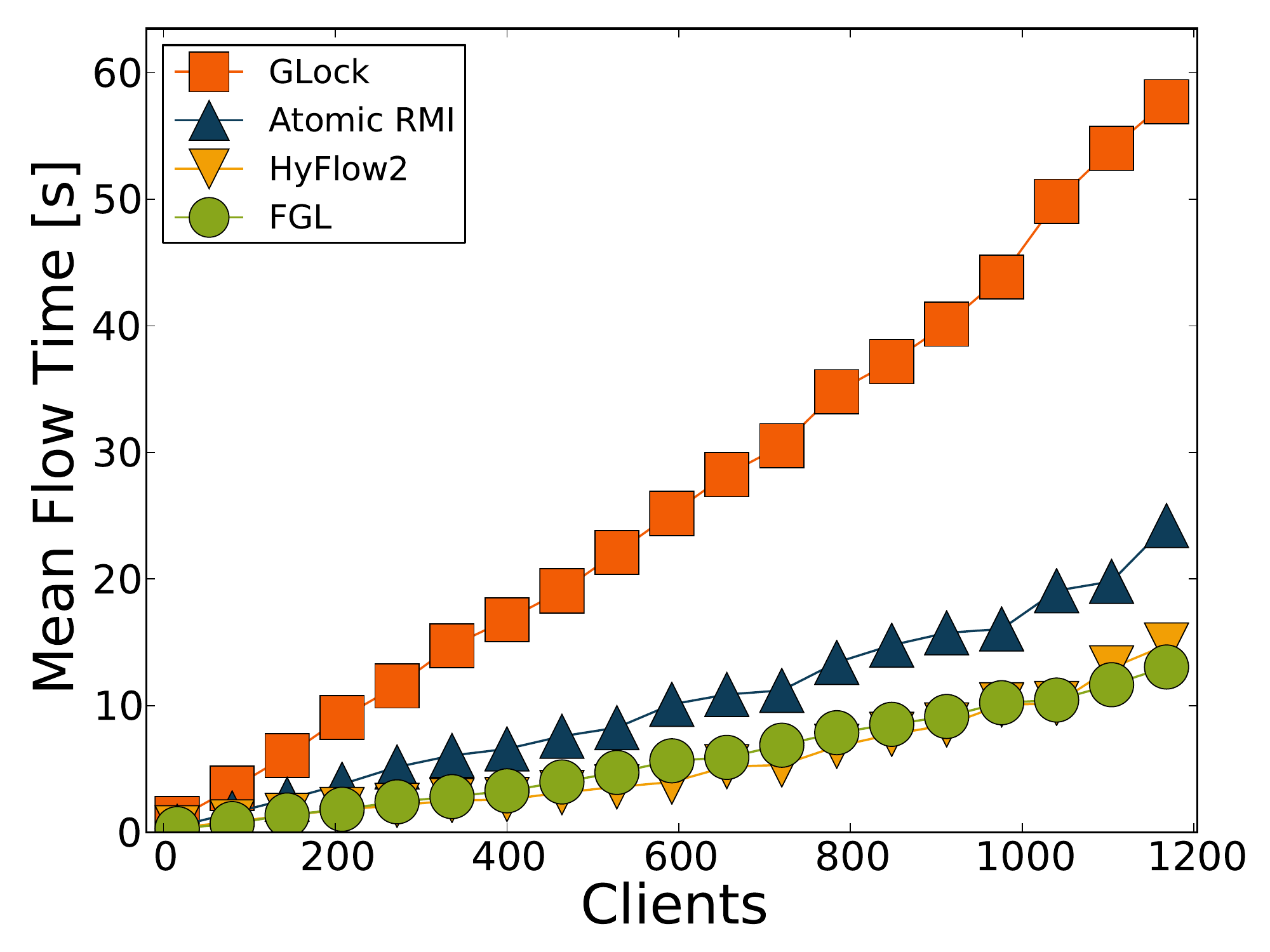}
\caption{\label{fig:mft}Mean flow time for the standard scenario.\vspace{\baselineskip}}
\end{minipage}
\end{figure}

The \emph{Large R} scenario (\rfig{fig:large-r}) shows similar behavior to
\emph{Small R}, but longer transactions have an impact on the performance of
all four frameworks, that is especially visible in the 1--400 client range when
the benchmark has relatively low contention.  Note that transactional
throughput values for all frameworks are nevertheless much lower than in
\emph{Small R}, since the transactions are much longer on average. At the
outset, FGL, HyFlow2, and Atomic RMI all achieve comparatively low throughput,
which increases as new clients are added. We attribute this to the initial
undersaturation of the system, which means that adding new clients only
increases tasks executed in parallel, without increasing the incidence of
conflicts. Hence, Atomic RMI gradually increases throughput until it reaches 50
transactions per second at the 400 client mark. On the other hand, the
throughput of HyFlow2 and FGL increases more rapidly with the introduction of
new clients and peaks with around 100 clients at 70--75 transactions per
second. This, however saturates both frameworks, and any further added clients
cause a decline in performance, until both frameworks stabilize around the 600
client mark at about 30--40 transactions per second. Hence, as in \emph{Small
R}, there is range of load for which optimistic TM or a more subtle locking
solution is preferable to pessimistic TM, but the distinction is much more
pronounced. On the other hand, once contention reaches a certain threshold,
pessimistic TM gains a stable advantage over the other two types of frameworks.
Thus, the choice of paradigm must be tailored to the application and its
workload, and \benchmark{} provides important clues as to how this adjustment
should be made.

When large read-write tasks are introduced in both \emph{Large W}
(\rfig{fig:large-w}) as well as \emph{Large R/W} (\rfig{fig:large-rw}), they
throttle the throughput of all frameworks, since they have long execution
times, as well as large R/W sets, and therefore typically cannot be executed in
parallel to other tasks. Hence, throughput is below 8 transactions per second
even for low contention situations. Furthermore, all frameworks perform
similarly to GLock with Atomic RMI performing marginally better, FGL performing
the same, and HyFlow2 performing notably worse. The performance of optimistic
TM below GLock is attributable to the tendency of the workload to generate a
large incidence of conflicts, leading to a high abort rate (52.5\% for
\emph{Large R/W} and 68.8\% for \emph{Large W}) and a hight retry rate (211
retries per commit in \emph{Large R/W} and 803 for \emph{Large W}).

\subsection{Aborts}

We show the abort and retry rates for \rfig{fig:aborts}. The figure shows the
results for HyFlow2, whereas the remaining frameworks had a consistent abort
rate of 0 (i.e. no aborts at all) and a retry rate of 1 (i.e. one attempt to
execute a transaction per commit). The results for HyFlow2 show that the abort
rate of that implementation tends to fluctuate around a certain stable level,
despite increasing contention with the number of clients. However, the retry
rate shows, that transactions that abort tend to abort more often on average as
the contention increases, leading to a high number of aborts total.

\subsection{Mean Flow Time}

The benchmark allows to perform other measurements, including mean flow time.
We show an example of that measurement in \rfig{fig:mft} for the standard
scenario. Mean flow time indicates, that whereas the analysis of throughputs
prefer Atomic RMI over FGL and HyFlow2 in the standard scenario, the average
time between a transaction's start and commitment is lower in FGL and HyFlow2.
Hence, for systems, where short response time is important (e.g. from the user
experience perspective), those systems are preferable.

\subsection{Discussion}

The evaluation shows that the benchmark provides a wide range of articulation
through its parameter selection and the composition of the workload. However,
despite the achieved configurability, the benchmark remains rooted in a
real-life application and reflects the complexity of the implementation of such
a system, which differenciates it from microbenchmarks.
This connection to reality makes the conclusions drawn on the basis of the
results more meaningful and applicable to practical use cases. 

For example, given a specific application like a distributed data warehouse for
use in analytics and decision support, where it is characteristic to expect
predominantly read-only transactions that aggregate data from various sources,
the evaluation data gathered in the evaluation indicates that in this particular use-case pessimistic TM
would perform better on average than other types of concurrency control schemes
(\rfig{fig:large-r}). However, we also note that this is the case only if the
contention is fairly high. To the contrary, as the number of clients depreciates, an
optimistic TM is the better choice to capitalize on low overhead.
This data is therefore immediatelly useful to an architect involved in the
design of such an application.

Another example is that of a geographically distributed system of sensors
periodically checked by a network of servers, that use transactions to maintain
consistency. Here, a large quantity of small read operations is expected as in
\rfig{fig:small-r}. The case is similar to the one above, but while optimistic
TM is significantly more performant in low contention, its throughput degrades
in high contention, while pessimistic TM performs in a more stable fashion.
This allows the architect to select either TM based on the variability of the
server network.

On the contrary to both previous system types, the data from \rfig{fig:large-w}
shows that when designing a datastore used primarily for archivization, i.e.
one with a large amount of long and write-oriented transactions, one need not
bother with sophisticated synchronization schemes, since the number of
conflicts involved forces near-serial execution in any case.

Alternatively, developers of TM systems can use the data provided by
\benchmark{} to analyze the systems and easily find their strong and weak
points, whether inherent or ammeanable. 
For instance, as presented above HyFlow2 shows some room for improvement when
subjected to write-heavy workloads with high retry rates. Knowing this, the
developers of HyFlow2 
can tune their contention management.
They can still use \benchmark{} for this purpose by creating
some specialised scenarios for detailed examination of very specific situations
and workloads. This can lead to the identification of the source of such
behavior in the TM design. As for Atomic RMI, its notably worse performance in
terms of MFT comparing to the HyFlow2 can be at least partially attributed to
the fact that this system does not distinguish between read and write
operations. Presumably introducing such distinction in the internal workings of
this system and maximizing the parallelization of read operations can lead to
closing the MFT performance gap between said TM systems.

%% file: ScenariosTable.tex
\begin{figure}[t]
{\footnotesize
\begin{center}
\begin{tabular}%
               {|l|c|c|c|c|c|c|c|}
\hline
Task type & Standard & Small R & Small R/W & Small W & Large R & Large R/W & Large W \\ \hline\hline
term search        & 0.25 & 0.30 & 0.19 & 0.02 & 0.44 & 0.19 & 0.02 \\
interaction search & 0.20 & 0.30 & 0.19 & 0.02 & 0.02 & 0.02 & 0.02 \\
send unicast       & 0.06 & 0.02 & 0.19 & 0.44 & 0.02 & 0.02 & 0.02 \\
send multicast     & 0.04 & 0.02 & 0.02 & 0.02 & 0.02 & 0.19 & 0.30 \\
batch import       & 0.04 & 0.02 & 0.02 & 0.02 & 0.02 & 0.18 & 0.30 \\
clear inbox        & 0.06 & 0.02 & 0.18 & 0.44 & 0.02 & 0.19 & 0.30 \\
association level  & 0.20 & 0.30 & 0.19 & 0.02 & 0.02 & 0.02 & 0.02 \\
indexing           & 0.15 & 0.02 & 0.02 & 0.02 & 0.44 & 0.19 & 0.02 \\
\hline
\end{tabular}
\end{center}
\caption{\label{fig:tasks}Task type probabilities in evaluation scenarios.}
}
\end{figure}

%% file: AbortRate_RetryRate.tex
\begin{tabular}
{|c|c|c|}
\hline
\# Clients & Abort ratio & Retry rate \\ \hline\hline
16 & 11.8\% & 1.255 \\
80 & 22.4\% & 2.900 \\
144 & 17.7\% & 4.516 \\
208 & 26.1\% & 6.044 \\
272 & 21.8\% & 7.744 \\
336 & 30.2\% & 8.625 \\
400 & 23.1\% & 10.278 \\
464 & 21.5\% & 12.903 \\
528 & 30.9\% & 13.909 \\
592 & 29.5\% & 16.805 \\
656 & 30.1\% & 18.522 \\
720 & 23.3\% & 20.657 \\
784 & 22.5\% & 23.772 \\
848 & 21.8\% & 26.812 \\
912 & 29.0\% & 27.750 \\
976 & 25.9\% & 31.424 \\
1040 & 26.0\% & 34.406 \\
1104 & 21.8\% & 36.263 \\
1168 & 26.8\% & 38.932 \\
\hline
\end{tabular}
\caption{\label{fig:aborts}HyFlow2 abort ratio and retry rate for the standard scenario.}

%% file: conclusions.tex
\section{Conclusions}
\label{sec:conclusions}

We presented \benchmark{}, a benchmark for performing comprehensive evaluations
and comparisons of distributed TM systems. It provides depth of analysis by its
implementation of a complex use case. It also allows for analyzing many aspects
of the memory using a wide variety of parameters and metrics. Finally, it is
based on a real application, meaning that the results of evaluations are more
likely to be reflected in practice when the TM system is deployed in the real
world. As such, it fills the vacuum caused by a lack of distributed TM
benchmarks other than microbenchmarks.

We showed that \benchmark{} provides important and meaningful data regarding TM
system performance in near real-life situation. \benchmark{}, in contrast to
other benchmarks that have been used for testing distributed TMs, is built
particularly for evaluating distributed TMs, and allows to do so with specific
distributed applications in mind.  Users of \benchmark{} can use several,
different parameters to create specialized scenarios to suit their specific
needs and use them to perform extenvie and complex evaluation.

In our future work we wish to create a full suite of benchmarks for testing
distributed TMs, to fulfill the requirement for broad evaluations, as well as
deep ones. That is, in order to be able to evaluate distributed TMs in
different applications from different fields. To that end we plan on distributing some
applications from the STAMP benchmark suite, as well as to find additional
applications that are specific to distributed systems. The work on these is
already underway and the early results are promising.

%% file: top-arxiv.bbl
\begin{thebibliography}{10}

\bibitem{AKJ+08}
M.~Ansari, C.~Kotselidis, K.~Jarvis, M.~Luján, C.~Kirkham, and I.~Watson.
\newblock {Lee-TM}: A non-trivial benchmark for transactional memory.
\newblock In {\em {Proceedings of~}ICA3PP'08: the 8th International Conference
  on Algorithms and Architectures for Parallel Processing}, June 2008.

\bibitem{ADE+01}
V.~Aslot, M.~Domeika, R.~Eigenmann, G.~Gaertner, W.~B. Jones, and B.~Parady.
\newblock {SPEComp}: A new benchmark suite for measuring parallel computer
  performance.
\newblock In {\em {Proceedings of~}WOMPAT'01: the International Workshop on
  OpenMP Applications and Tools: OpenMP Shared Memory Parallel Programming},
  July 2001.

\bibitem{BAC08}
R.~L. Bocchino, V.~S. Adve, and B.~L. Chamberlain.
\newblock {Software Transactional Memory for Large Scale Clusters}.
\newblock In {\em {Proceedings of~}PPoPP'08: the 13th~{ACM SIGPLAN Symposium on
  Principles and Practice of Parallel Programming}}, Feb. 2008.

\bibitem{COR12}
J.~C. Corbett and et~al.
\newblock {Spanner: {G}oogle's Globally-Distributed Database}.
\newblock In {\em {Proceedings of~}OSDI'12: the 10th~{USENIX Symposium on
  Operating Systems Design and Implementation}}, Oct. 2012.

\bibitem{CRCR09}
M.~Couceiro, P.~Romano, N.~Carvalho, and L.~Rodrigues.
\newblock {D2STM: Dependable Distributed Software Transactional Memory}.
\newblock In {\em {Proceedings of~}{PRDC'13: the 15th IEEE Pacific Rim
  International Symposium on Dependable Computing}}, Nov. 2009.

\bibitem{GK10}
R.~Guerraoui and M.~Kapa{\l}ka.
\newblock {\em {Principles of Transactional Memory}}.
\newblock Morgan \& Claypool, 2010.

\bibitem{GKV07}
R.~Guerraoui, M.~Kapa{\l}ka, and J.~Vitek.
\newblock {STMBench7}: A benchmark for software transactional memory.
\newblock In {\em {Proceedings of~}EuroSys'07: the 2nd~{ACM SIGOPS/EuroSys
  European Conference on Computer Systems}}, June 2007.

\bibitem{HM93}
M.~Herlihy and J.~E.~B. Moss.
\newblock {Transactional Memory: Architectural Support for Lock-free Data
  Structures}.
\newblock In {\em {Proceedings of~}ISCA'93: the 20th~{International Symposium
  on Computer Architecture}}, pages 289--300, May 1993.

\bibitem{HPR14}
S.~Hirve, R.~Palmieri, and B.~Ravindran.
\newblock {HiperTM: High Performance, Fault-Tolerant Transactional Memory}.
\newblock In {\em {Proceedings of~}{ICDCN'14: the 15th~}{International
  Conference on Distributed Computing and Networking}}, Jan. 2014.

\bibitem{HOCB+10}
S.~Hong, T.~Oguntebi, J.~Casper, N.~Bronson, C.~Kozyrakis, and K.~Olukotun.
\newblock Eigenbench: A simple exploration tool for orthogonal tm
  characteristics.
\newblock In {\em {Proceedings of~}IISWC'10: the IEEE International Symposium
  on Workload Characterization}, 2010.

\bibitem{KKW13}
T.~Kobus, M.~Kokoci\'nski, and P.~T. Wojciechowski.
\newblock Hybrid replication: State-machine-based and deferred-update
  replication schemes combined.
\newblock In {\em {Proceedings of~}{ICDCS'13: the 33rd~}{International
  Conference on Distributed Computing Systems}}, July 2013.

\bibitem{LM10}
A.~Lakshman and P.~Malik.
\newblock Cassandra: a decentralized structured storage system.
\newblock {\em {ACM SIGOPS Operating Systems Review}}, 44:25--40, Apr. 2010.

\bibitem{MCKO08}
C.~C. Minh, J.~Chung, C.~Kozyrakis, and K.~Olukotun.
\newblock {STAMP: Stanford Transactional Applications for Multi-Processing}.
\newblock In {\em {Proceedings of~}IISWC'08: the IEEE International Symposium
  on Workload Characterization}, Sept. 2008.

\bibitem{MTPR13}
S.~Mishra, A.~Turcu, R.~Palmieri, and B.~Ravindran.
\newblock {HyflowCPP}: A distributed transactional memory framework for {C++},
  Aug. 2013.

\bibitem{PD10}
D.~Peng and F.~Dabek.
\newblock Large-scale incremental processing using distributed transactions and
  notifications.
\newblock In {\em {Proceedings of~}OSDI '10: 9th USENIX Symposium on Operating
  Systems Design and Implementation}, Oct. 2010.

\bibitem{RLS14}
W.~Ruan, Y.~Liu, and M.~Spear.
\newblock Stamp need not be considered harmful.
\newblock In {\em {Proceedings of~}TRANSACT'14: the 9th~{ACM SIGPLAN Workshop
  on Transactional Computing}}, Feb. 2014.

\bibitem{SR12}
M.~M. Saad and R.~B.
\newblock Transactional forwarding: Supporting highly-concurrent {STM} in
  asynchronous distributed systems.
\newblock In {\em {Proceedings of~}SBAC-PAD'12: the 24th IEEE International
  Symposium on Computer Architecture and High Performance Computing}, Oct.
  2012.

\bibitem{SR11}
M.~M. Saad and B.~Ravindran.
\newblock {HyFlow: A High Performance Distributed Transactional Memory
  Framework}.
\newblock In {\em {Proceedings of~}{HPDC '11: the 20th International Symposium
  on High Performance Distributed Computing}}, June 2011.

\bibitem{SW14-hlpp}
K.~Siek and P.~T. Wojciechowski.
\newblock {Atomic RMI: a Distributed Transactional Memory Framework}.
\newblock In {\em {Proceedings of~}{HLPP'14: the 7th International Symposium on
  High-level Parallel Programming and Applications}}, pages 73--94, July 2014.

\bibitem{SW14-disc}
K.~Siek and P.~T. Wojciechowski.
\newblock Brief announcement: Relaxing opacity in pessimistic transactional
  memory.
\newblock In {\em {Proceedings of~}DISC'14: the 28th~{International Symposium
  on Distributed Computing}}, 2014.

\bibitem{SW15-ijpp}
K.~Siek and P.~T. Wojciechowski.
\newblock {Atomic RMI: A Distributed Transactional Memory Framework}.
\newblock {\em International Journal of Parallel Programming}, 2015.

\bibitem{TRP13}
A.~Turcu, B.~Ravindran, and R.~Palmieri.
\newblock {HyFlow2: A High Performance Distributed Transactional Memory
  Framework in Scala}.
\newblock In {\em {Proceedings of~}{PPPJ'13: the 10th~International Conference
  on Principles and Practices of Programming on JAVA platform: virtual
  machines, languages, and tools}}, Sept. 2013.

\bibitem{Woj07}
P.~T. Wojciechowski.
\newblock {\em {Language Design for Atomicity, Declarative Synchronization, and
  Dynamic Update in Communicating Systems}}.
\newblock Pozna{\'n} University of Technology Press, 2007.

\bibitem{WOT+95}
S.~C. Woo, M.~Ohara, E.~Torrie, J.~P. Singh, and A.~Gupta.
\newblock The {SPLASH-2} programs: characterization and methodological
  considerations.
\newblock In {\em {Proceedings of~}ISCA'95: the 22nd Annual International
  Symposium on Computer Architecture}, May 1995.

\bibitem{ZGU+09}
F.~Zyulkyarov, V.~Gajinov, O.~S. Unsal, A.~Cristal, E.~Ayguad{\'e}, T.~Harris,
  and M.~Valero.
\newblock {Atomic Quake}: using transactional memory in an interactive
  multiplayer game server.
\newblock In {\em {Proceedings of~}PPoPP'09: the 14th~{ACM SIGPLAN Symposium on
  Principles and Practice of Parallel Programming}}, Apr. 2009.

\end{thebibliography}
